\documentstyle[12pt]{article}   

\large
\newcommand{\tiN}{\raisebox{-6.5pt}{$\displaystyle
\stackrel{\displaystyle N}{\sim}$}}

\newcommand{\beq}{\begin{equation}}
\newcommand{\eeq}{\end{equation}}

\makeatletter
\@addtoreset{equation}{section}
\makeatother

\input epsf

\title{The reduced phase space of spherically symmetric Einstein-Maxwell
       theory including a cosmological constant}
\author{T. Thiemann\thanks{present address : Center for Gravitational Physics
and Geometry, The Pennsylvania State University, University Park,
PA 16802-6300, USA} 
\thanks{thiemann@phys.psu.edu} \\
       Institute for Theoretical Physics, RWTH Aachen,\\
       52056 Aachen, Germany}
\date{{\small Preprint PITHA 93-32, August 93}}

\begin{document}

\maketitle                     

\begin{abstract}

We extend here the canonical treatment of spherically symmetric (quantum)
gravity to the most simple matter coupling, namely               
spherically symmetric Maxwell theory with or without a cosmological constant.
The quantization is based on the reduced phase space which is coordinatized
by the mass and the electric charge as well as their canonically conjugate
momenta, whose geometrical interpretation is explored.\\
The dimension of the reduced phase space depends on the topology chosen,
quite similar to the case of pure (2+1) gravity.\\
We investigate several conceptual and technical details that might be of
interest for full (3+1) gravity.
We use the new canonical variables introduced by Ashtekar,
which simplifies the analysis tremendously.

\end{abstract}

\section{Introduction}

The introduction of a new set of canonical variables due to Ashtekar
\cite{4} brings the initial value constraints of general 
relativity into {\em polynomial} form. This tremendous simplification of the 
algebraic form of the constraint functionals has far reaching 
consequences : since the first step in the canonical quantization
programme is the solution of the constraints (meaning that an appropriate
operator version of the constraint functionals vanishes on physical
states) there is justified hope that with these new canonical
variables the roadblock that one meets when using the old (ADM) \cite{15} 
variables (namely that the constraints are not even analytical in 
the basic variables so that it is not even clear how to define the 
constraint operators) can actually be overcome.

It is suggested to verify this assumption by first trying to quantize
simplified models of pure and matter-coupled gravity. As expected,
the complete quantization of model systems in the new variables 
(e.g. \cite{1,2,5,16,26}) proves to be more feasible and actually led to 
new, 
surprising results. Moreover, it gives some confidence in attacking the
much more complicated problem of quantizing full (3+1) general relativity
via canonical techniques.

Up to now, no model for gravity with matter has been quantized in the new
variables. Especially interesting would be a gauge field as the matter 
content 
because in this case the scalar constraint turns out to be of {\em fourth},
rather than second, order in the momenta which is a quite unusual situation 
that has been barely dealt with in the literature so far, in fact, the 
author is not aware of any such case.

In this paper we are going to rigorously quantize a spherically 
symmetric charged (Reissner-Nordstr{\o}m \cite{15}) black hole. 
This is the most simple gauge 
field that one can imagine to couple to gravity, simple enough in order that
all the steps of the canonical quantization programme can be carried out.
Only in that way one can hope to gain some insight in the problems that
will occur in the full theory and how to solve them. Although this model
only has a finite number of degrees of freedom after solving the constraints,
it is a true field theory to start with and therefore "more similar" to
full general relativity than, say, cosmological models.\\
We can also include a cosmological constant into the analysis if we choose
a closed (rather than asymptotically flat) topology for the initial value
hypersurface.

This paper is the natural extension of \cite{1} (hereafter referred to as
I). These two papers treat the vacuum case of a spherically symmetric 
black hole (similar results were obtained also in \cite{25}). Therefore,
this paper, although in principle self-contained, uses many
of the results (that is, computational results) of I in order to avoid 
repetition.

It turns out that the application of the operator constraint method
(Dirac method \cite{19}) fails due to the fourth order (in the momenta)
of the scalar constraint when using the Ashtekar variables. More
precisely, the author was unable to find consistent orderings (without
regularization) of the constraint operators such that they form a 
closed algebra (point splitting regularizations are available but the 
associated manipulations are ill-defined \cite{6}). However,
we succeed in using the framework of symplectic reduction (degenerate
Hamilton-Jacobi method \cite{10}).

The computational part of this paper is quite large, however, as it is 
well known, one actually looses {\em physical} degrees of freedom if
one is not very careful with the functional analysis, at least for 
asymptotically flat topologies. It is therefore for physical reasons that
we display in detail the fine tuned interplay between the role of 
boundary conditions (fall-off behaviour
of the various fields), the difference between symmetries and gauge, the
finiteness of the functionals on the reduced phase space and their
(functional) differentiability. These items show up only in the symplectic
reduction of {\em true} field theories, models that are finite-dimensional
even before reduction like cosmologies (see ref. \cite{5}) or those that
deal only with closed topologies like (2+1) gravity (see ref. \cite{2}) do
not share these problems and so it seems to be justified to dwell a little
bit on the mathematical techniques involved.\\
\\
The organization of the paper is as follows :\\
\\
Section 2 introduces the model in terms of Ashtekar's variables. This serves
mainly in order to fix our notation.

Section 3 discusses the various available topologies for the initial data
hypersurface. After a topology is
fixed (once and for all - this is one of the disadvantages of the canonical
quantization scheme), the associated fall-off behaviour of the various
fields can be derived. It will turn out that the dimension of the reduced
phase space depends on the topology chosen - quite similar to the case of
the (2+1) gravity model. The role of the genus of the Riemann surface which
is the initial value surface for (2+1) gravity is played by the number of
the asymptotic ends for spherically symmetric gravity (irrespective of
whether with or without sources).\\
The fall-off behaviour of the fields is chosen more general than the one
used by Ashtekar \cite{4}.

Section 4 then comprises the symplectic reduction of the present model.
Quite surprisingly, all constraints can be strictly solved for the momenta
which consist of the Einstein-Maxwell 'electric' fields in the Ashtekar
polarization of the phase space. However, the equations for the electric
fields become 4th order algebraic equations and hence we were not able
to apply the theorem proved in I in order to solve the associated
Hamilton-Jacobi equations due to the complicated appearance of the
solution formulas for the electric fields. The basic idea is to change the
polarization in order to simplify the Hamilton-Jacobi equations and to
return to the Ashtekar-polarization after the solution has been found.

Section 5 discusses the topology and reality structure of the reduced
phase space obtained in section 4. The topology turns out to 
be quite complicated due to the appearance of several 
"sectors" (the solution of an n-th order scalar constraint in 
terms of its momenta has n roots) especially for non-zero 
cosmological constant. Furthermore we prove that with the
function spaces derived in section 3, the observables (i.e. coordinates on
the reduced phase space) are finite and functionally differentiable.

Having obtained this result, we are able to prove in section 6
that the observables are
gauge invariant but transform non-trivially under symmetries of the
asymptotic structure. This in turn allows to give a definite interpretation
of the observables conjugate to the mass and the electric charge
respectively, at least for open topologies : they are the eigentime of an
asymptotic observer and the variable conjugate to the electric charge that
also appears in canonical (1+1) Maxwell theory. The latter can be interpreted
as a 'formal magnetic flux' (which has nothing to do with the monopole that
we might or might not introduce) and we will refer to it as the magnetic flux
in the sequel. We want to stress that
these observables are genuine {\em volume} integrals and no surface terms
as has been expected by some authors 
which is important because otherwise the reduced
phase space for closed topologies would only carry a presymplectic, rather
than a symplectic structure. They do not seem to have been discovered 
before, probably because the Birkhoff theorem (\cite{15}) excludes the
existence of these variables. We recall from I why this is no contradiction.\\
Moreover, as already stressed in the second paper of \cite{1},
the reduced system
adopts the form of an integrable model whereby the role of the action
variables is played by the mass and the charge whereas the angle variables
are their canonically conjugate momenta.

Section 7 is dedicated to quantum theory.
After the quantum theory via the group theoretical approach is derived
(\cite{8}) one is able to study the solutions of the
Schroedinger-equation. The eigenvectors of the Hamilton-operator are peaked
on the classical eigentime and classical magnetic flux.\\
Another subsection deals with the operator constraint
or Dirac algorithm for quantizing field theories with constraints. As for
all Yang-Mills theories, the scalar constraint functional becomes
non-analytic in the electric fields so that this method (in the
Ashtekar-polarization) only makes sense after multiplying the scalar
constraint functional with an
appropriate power of the electric fields. The unregulated constraint operator
does not close the quantum algebra whereas the regulated one is ill-defined
as already said above.
If one works, however, with the polarization suggested already by the
classical theory, then one recovers the same quantum theory as obtained via
the reduced phase space method. 

The paper concludes with some remarks on what has been learnt by studying
this model.

In an appendix we carry out a tedious computation which proves the statements 
given in section 6.

\section{Introduction of the model}

We use the same notation as in I. The spherically symmetric reduction of the
Einstein sector of the model is defined identically as in I. Furthermore,
only gravitational fields contribute to the (ADM)-energy-momentum. Finally,
the reality conditions on the Ashtekar variables are unchanged when coupling
bosonic matter only, hence the formulas of I {\em except for the boundary
conditions} can be taken over without change directly to the present case.
Suffice it to recall that after the spherically symmetric reduction the 
Ashtekar variables $E^a_i=\det(e_b^j)e^a_i\mbox{ and 
}A_a^i=\Gamma_a^i+iK_{ab}e^b_i$, where $e^a_i$ is the spatial triad and
$K_{ab}$ is the extrinsic curvature of the initial data hypersurface,
become
\begin{eqnarray*}
&& (E^x_i,E^\theta_i,E^\phi_i)=(\sin(\theta)E^1 
n^x_i,\frac{\sin(\theta)}{\sqrt{2}}(E^2 n^\theta_i+E^3 n^\phi_i),
\frac{1}{\sqrt{2}}(E^2 n^\phi_i-E^3 n^\theta_i))\\
&& (A_x^i,A_\theta^i,A_\phi^i)=(A_1 n^x_i,\frac{1}{\sqrt{2}}(A_2 n^\theta_i
+(A_3-\sqrt{2}) n^\phi_i),\frac{\sin(\theta)}{\sqrt{2}}(A_2 n^\phi_i
-(A_3-\sqrt{2})n^\theta_i)) \;.
\end{eqnarray*}
Here $\theta,\phi$ are the usual polar coordinates on $S^2$ and the 
internal vectors $n_x,n_\theta,n_\phi$ are the standard orthonormal
vectors \cite{1}.
The functions $E^I,A_I,\; I=1,2,3$ depend on the spatial variable x and the
time variable t only. $E^I$ is real while $A_I-\Gamma_I$ is imaginary
where 
\[ (\Gamma_1,\Gamma_2,\Gamma_3)=(-\frac{(E^3)' E^2-(E^2)' 
E^3}{(E^2)^2+(E^3)^2},-\frac{(E^1)' E^3}{(E^2)^2+(E^3)^2},
\frac{(E^1)' E^2}{(E^2)^2+(E^3)^2}) \] 
We can now proceed to the source terms.\\
We require that the Maxwell electric ($\epsilon^a$) and magnetic fields
($\mu^a$) are spherically
symmetric, i.e. they are Lie annihilated by the generators of the SO(3)
Killing group. The unique solution of
this definition are radially symmetric fields
\begin{eqnarray}
(\epsilon^x,\epsilon^\theta,\epsilon^\phi) &:=& (\epsilon(x,t),0,0),\nonumber\\
(\mu^x,\mu^\theta,\mu^\phi) &:=& (\mu(x,t),0,0) .
\end{eqnarray}
Now we exploit that the magnetic field comes from a spatial potential
$\omega_a$ i.e. $\mu^a=1/2\epsilon^{abc}\partial_b \omega_c$. Then it follows
(locally) from the Bianchi identity
\beq 2\partial_x\mu^x=-(\partial_\theta\mu^\theta+\partial_\phi\mu^\phi)=0
\; ,\eeq
i.e. $\mu=\mu(t)$ is a spatial constant, the magnetic charge. The Maxwell
potential is thus given by
\beq (\omega_x,\omega_\theta,\omega_\phi)=(\omega(x,t),0,0)
+(\Omega_a(x,t,\theta,\phi)), \eeq
where $\Omega_a$ is a monopole solution with charge $\mu$ :
$\star d\wedge\Omega=\mu$ (it has no radial part). \\
The cosmological constant will be labelled by the (real) parameter $\lambda$.
Then we are already in the position to complete the reduction to spherical
symmetry by plugging the formulas (2.1) and (2.3) into the canonical 
Yang-Mills action (given below for arbitrary (semisimple) gauge group G)
\beq
^{YM}S=\int_R dt\int_{\Sigma^3} d^3x tr\{\dot{\omega}_a\epsilon^a-[-U{\cal D}_a
\epsilon^a+N^a\epsilon_{abc}\mu^c\epsilon^b+\tiN\frac{1}{2}q_{ab}
(\epsilon^a\epsilon^b+\mu^a\mu^b)]\}
\eeq
which for spherical symmetry and $G=U1)$ becomes after integration over the
sphere (using $\Sigma^3=S^2\times\Sigma\;\mbox{where}\;\Sigma$ is 
1-dimensional)
\beq
^M S=4\pi\int_R dt\int_\Sigma dx \{\dot{\omega}\epsilon-[-U\epsilon'
+\tiN\frac{1}{2}\frac{(E^2)^2+(E^3)^2}{2E^1}((\epsilon)^2+(\mu)^2)]\} \;.
\eeq
Note that the monopole potential was projected out since it has no radial
component and that for the same reason there is no Maxwell-contribution to
the vector constraint. Furthermore we do not care about boundary terms at
this stage.
We define $p^2:=\epsilon^2+\mu^2$, rescale the
Lagrange-multiplier U and the radial part of the Maxwell-connection by
$p/\sqrt{p^2-\mu^2}$ and we arrive at the same action without magnetic
charge. Formally we have carried out a 'duality rotation' which leaves the
energy-momentum tensor of the Maxwell-field unchanged.\\
Finally, we have for the cosmological constant term
\beq ^C S=\int_R dt\int_\Sigma d^3x N\lambda\sqrt{q}=4\pi \int_R dt
\int_\Sigma dx \lambda\tiN\frac{1}{2}((E^2)^2+(E^3)^2)E^1\; . \eeq
The model has thus 4 canonical pairs $(\omega,p\; ;\; A_I,E^I)$ and is
subject to the 4 constraints, defined by the following 4 constraint
functionals :\\
Maxwell Gauss constraint :
\beq ^M{\cal G}=p', \eeq
Einstein Gauss constraint :
\beq ^E{\cal G}=(E^1)'+A_2 E^3-A_3 E^2, \eeq
Vector constraint :
\beq V=B^2 E^3-B^3 E^2, \eeq
scalar constraint :
\beq C=(B^2 E^2+B^3 E^3)E^1+\frac{1}{2}((E^2)^2+(E^3)^2)(B^1
+\kappa\frac{p^2}{2 E^1}+\kappa\lambda E^1)
\eeq
which implies that there will be only a finite number of degrees of freedom
left on the reduced phase space.\\
Additionally to the ADM energy-momentum (I) there is the electric charge
boundary term (this interpretation follows from inserting a
Reissner-Nordstr{\o}m solution)
\beq +\int_{\partial\Sigma}U p \; .\eeq

\section{Topologies and function spaces}

\subsection{Possible topologies}

For spherically symmetric systems, the topology of the 3 manifold is
necessarily of the form $\Sigma^{(3)}=S^2\times\Sigma$ where $\Sigma$ is a
1-dimensional manifold.\\
We will deal with 2 kinds of topologies :\\
a) compact without boundary \\
In this case the only possible choice is $\Sigma=S^1$, i.e. we have the
topology of a compactified wormwhole.\\
b) open with boundary\\
As was motivated already in I, we choose now
\[ \Sigma=\Sigma_n \; ,\; \Sigma_n\cong K\cup\bigcup_{A=1}^n\Sigma_A \; ,
\]
i.e. the hypersurface is the union of a compact set K (diffeomorphic to a
compact interval) and a collection of ends (each of which is diffeomorphic
to the positive real line without the origin) i.e. asymptotic regions
with outward orientation and all of them are joined to K. This means, we
have n positive real lines, including the origin, but one end of each line
is common to all of them, i.e. these parts are identified. Since the
identity map is smooth, this is still a $C^\infty$ (Hausdorff) manifold
except in a neighbourhood of the end point of the common line.
This kind of topology is illustrated in the figure below.\\
\unitlength 1cm
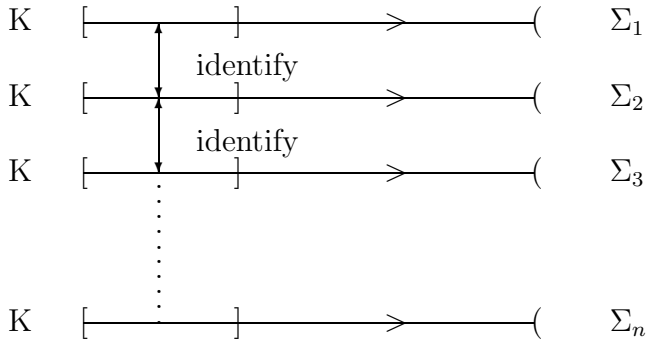
\begin{picture}(9,6)
\put(0.95,4.9){[}\put(1,5){\line(6,0){6}}\put(3,4.9){]}\put(6.95,4.9){(}
\put(5,4.9){$>$}\put(8,4.9){$\Sigma_1$}
\put(0.95,3.9){[}\put(1,4){\line(6,0){6}}\put(3,3.9){]}\put(6.95,3.9){(}
\put(5,3.9){$>$}\put(8,3.9){$\Sigma_2$}
\put(0.95,2.9){[}\put(1,3){\line(6,0){6}}\put(3,2.9){]}\put(6.95,2.9){(}
\put(5,2.9){$>$}\put(8,2.9){$\Sigma_3$}
\put(0.95,0.9){[}\put(1,1){\line(6,0){6}}\put(3,0.9){]}\put(6.95,0.9){(}
\put(5,0.9){$>$}\put(8,0.9){$\Sigma_n$}
\put(2,4){\vector(0,1){1}}\put(2,5){\vector(0,-1){1}}
\put(2,3){\vector(0,1){1}}\put(2,4){\vector(0,-1){1}}
\put(2.5,4.3){identify}\put(2.5,3.3){identify}
\multiput(1.95,1)(0,0.2){10}{.}
\put(0,4.9){K}\put(0,3.9){K}\put(0,2.9){K}\put(0,0.9){K}
\end{picture}
\\
We want to point out here three items :\\
1) The boundary of the compactum K is to be understood topologically,
for a given value of the mass m it is not fixed a priori at the coordinate
value $r=m$ because otherwise the topology would depend on a dynamical
object, the mass of the system, while in the canonical framework the 
choice of the topology is a kinematical ingredient, it is fixed once and for
all right from the beginning before solving the dynamics of the system.\\ 
2) Boundary conditions should not only be imposed in the asymptotic regions
but also at the origin inside the compactum K. Note that this is never done
in the literature since one is usually only interested in issues
like the positivity of the gravitational energy (\cite{20}) or the asymptotic
Poincar\'e group (\cite{9}).\\
3) The compactum K could be replaced by another topological object so as to
obtain the Reissner-Nordstr{\o}m topology (\cite{15}) that avoids closed 
timelike 
curves. The formalism does not force us to do that, that is,
the Reissner-Nordstr{\o}m topology does not follow from Einstein's equations
because the then necessary extension of K lies outside the domain of 
dependence of the initial data hypersurface. This is also why we do not have
to distinguish between the cases where the electric charge squared is less
or greater than or equal to the mass squared of the black hole. 
We refrain from analyzing the most general situation in the present paper 
and prefer to deal with a geodesically incomplete manifold.

\subsection{Derivation of the function spaces}

a) Asymptotically flat topologies\\
\\
Up to now we did not modify the Einstein sector of the theory at all compared
to I.\\ 
In contrast to I we will derive a new set of function spaces (fall-off 
behaviour of the fields) based
on the following 2 {\em minimal} requirements, following \cite{9} : \\
1) finiteness of the symplectic structure, \\
2) finiteness and functional differentiability of the constraint functionals.
Requirement 2) further depends on the set of asymptotic symmetries that one
is willing to allow.\\
In \cite{9} (which is based on the old (ADM) variables) these requirements
1) and 2) including asymptotical Poincare transformations can be satisfied as
follows :\\
\begin{eqnarray}
q_{ab}&\rightarrow& ^0 q_ab+\frac{f_{ab}(x^c/r,t)}{r}+O(1/r^{1+\epsilon}) 
\nonumber\\
p^{ab}&\rightarrow& \frac{k^{ab}(x^c/r,t)}{r^2}+O(1/r^{2+\epsilon})
\end{eqnarray}
as $r\rightarrow\infty, \; r:=^0 q_{ab} x^a x^b$ whereby $^0 q_{ab}$ is a fixed
(nondynamical) flat metric of Euclidean signature and $x^a$ are cartesian
coordinates with respect to it. Furthermore, it must be
required that the functions $f_{ab}$ and $k^{ab}$ respectively are even and
odd respectively under reflections of the asymptotically flat frame.\\
It is clear that for spherical symmetry we are not able to impose the above
parity conditions because the reduction to spherical symmetry excludes
all modes of the fields (regarded as expanded into spherical harmonics)
which have angular momentum different from zero. Hence we have to modify
the strategy slightly.\\
Comparing the spherically symmetric metric
\beq q_{ab}=\frac{(E^2)^2+(E^3)^2}{2E^1}x_{,a}x_{,b}+E^1 h_{ab} \eeq
with the Euclidean metric in spherical coordinates (we choose the coordinate
x to coincide asymptotically with the radial coordinate of the asymptotical
Euclidean frame)
\beq ^0 q_{ab}=x_{,a}x_{,b}+x^2 h_{ab} \; , \eeq
where $h_{ab}$ is the standard metric on the sphere, we conclude the
following fall-off\\ properties :
\begin{eqnarray}
 & & (E^1,E^2,E^3)\rightarrow (x^2[1+\frac{f^1(t)}{x}+O(1/x^2)], \nonumber\\
 & & \sqrt{2}x[\bar{E}^2+
\frac{f^2(t)}{x}+O(1/r^2)],\sqrt{2}x[\bar{E}^3+\frac{f^2(t)}{x}+O(1/x^2)])
\end{eqnarray}
whereby $(\bar{E}^2)^2+(\bar{E}^3)^2=1$. As well as the motivation for the
fall-off behaviour of the metric is that it should approach asymptotically a
Schwarzschild solution, the motivation for the fall-off of the Maxwell
electric field is that it should approach asymptotically a Coulomb solution.
The Coulomb solution in Minkowski space in spherical coordinates is just
$\epsilon^a=e x_{,a}$ where $e_{,x}=0$, hence we conclude
\beq p\rightarrow e(t)+O(1/x). \eeq
The Einstein-Maxwell connection behaves as a one-form under diffeomorphisms.
That means that compared to cartesian coordinates in spherical coordinates
the $\theta,\phi$ components adopt an additional power of x. Recalling
the definition of the Ashtekar connection from \cite{4} as well as the
definition of the Maxwell electric field $\epsilon^a=q^{ab}(F_{tb}
+N^c F_{bc})/\tiN$ which reduces here to $p=q^{xx}/\tiN(\dot{\omega}-U')$ we
conclude that
\begin{eqnarray}
(A_1,A_2,A_3-\sqrt{2}) & \rightarrow &(\frac{a_1(t)}{x^2}+O(1/x^3),
\frac{a_2(t)}{x}+O(1/x^2),\frac{a_3(t)}{x}+O(1/x^2)) \nonumber \\
\omega & \rightarrow & \frac{b(t)}{x^2}+O(1/x^3) \; .
\end{eqnarray}
Since, as we noted before, there is no parity freedom left, the requirements
1) and 2) discussed above will not be satisfied yet. Let us explore what
further restrictions are there to be imposed.\\
The symplectic structure on the large phase space can be read off from the
action. The non-vanishing brackets are
\beq \{A_I(x),E^J(y)\}=i\kappa\delta^J_I\delta(x,y) \;,\; \{\omega(x),p(y)\}
=\delta(x,y) \;\mbox{for all x,y in}\; \Sigma \; . \eeq
Written as a 2-form on the space of the variations of the fields :
\begin{eqnarray}
& &  \Omega=\int_\Sigma dx [-i/\kappa dE^I\wedge dA_I+dp\wedge d\omega]
\nonumber\\
& &=\int_\Sigma dx (\frac{-i}{\kappa x}[da_1\wedge df^1+\sqrt{2}(da_2\wedge
df^2+da_3\wedge df^3)]+O(1/x^2)) \; .
\end{eqnarray}
Hence we can satisfy requirement 1) by restricting the variations to be such
that
\beq da_1\wedge df^1+\sqrt{2}(da_2\wedge df^2+da_3\wedge df^3)=0 .\eeq
As for requirement 2) we first have to agree on the set of allowed symmetries
at infinity. We want to incorporate only asymptotic translations
as well as asymptotic U(1)-transformations of the Maxwell-field, and do also
allow for asymptotic O(2) transformations of the Einstein fields. Why do we
not consider asymptotic boosts of the 2-dimensional flat structure (rotations
do not exist in 1 dimension anyway) ? In the literature, one looks at
Schwarzschild-solutions in arbitrarily boosted frames (see ref. \cite{9}, for
example). However, these boosts are really boosts with respect to the
4-dimensional spacetime which violate spherical symmetry of the initial data.
The 'boosts' that we were able to discuss here must be meant with respect to
the effective 2-dimensional spacetime coordinatized by the variables x and t
in order not to violate spherical symmetry, they are thus not physical
anyway. But since we do not have this parity freedom at our disposal our
'boost' generator diverges. So we would have to impose much more
restrictive fall-off conditions than in (3.4)-(3.6) which, in particular, would
exclude Reissner-Nordstr{\o}m configurations and for that reason we refrain
from doing so. An option would be to impose some 'reflection conditions'
in different ends of $\Sigma$ to make the boost generator converge, however
then masses and charges in {\em different} ends would not evolve
independently of each other although they are spacelike separated and this
seems to contradict causality. Hence we neither impose such a condition.\\
The same is actually true for asymptotic translations : only radial
translations preserve the spherical symmetry of the fields, that is,
translations of the form $x^a\rightarrow x^a+c x^a/r$ where c is a constant
but these are then position-dependent (on the sphere) and do not correspond
to the translation subgroup of the Poincar\'e group, rather they are odd
supertranslations (\cite{9}). They correspond to a
translation of the radial coordinate r by c. Recalling that
$N^x=N^a\partial x/\partial x^a$ we have in this case (i.e. for
$N^a=cx^a/r$) really $N^x=c$.\\
Obviously, we have then {\em for symmetry transformations} the following
fall-off behaviour of the Lagrange multipliers :
\beq
(\Lambda,N^x,\tiN,U)\rightarrow(\frac{\mbox{const.}}{x^2}+O(1/x^3),
\mbox{const.}
+O(1/x),\frac{\mbox{const.}}{x^2}+O(1/x^3),\mbox{const.}+O(1/x))
\eeq
while {\em for gauge transformations} we require, for simplicity, that the
Lagrange multipliers are of compact support.\\
We now compute the leading order behaviour of the integrands of the
constraint functionals : the Maxwell Gauss constraint functional is already
finite, it becomes functionally differentiable when adding the electric
charge counterterm. For the rest of the constraints we have
\beq ^E{\cal G}\rightarrow 2x(1-\bar{E}^2)+f^1+\sqrt{2}(a_2\bar{E}^3-
\sqrt{2}f^2-\bar{E}^3 a_3)+O(1/x) \eeq
which becomes a finite functional when imposing $\bar{E}^2=1$ i.e. $\bar{E}^3
=0$.
Functional differentiability can be achieved without adding the O(2) charge
given in I.\\
We want here to draw attention to the following subtlety : the constraints
follow from setting the variation of the action with respect to the Lagrange
multipliers equal to zero. If now the variation of a Lagrange multiplier
happens to occur outside its support, its variation also vanishes. Hence
the constraints hold only in the support of the Lagrange multipliers.
What support is valid : that for symmetries or for gauge ? Since the
constraint equations are field equations the variations are set equal to
zero at spacelike and timelike infinity upon deriving the Euler-Lagrange
equations, hence it is consistent to impose the constraints
only off the boundaries although for simplicity one usually imposes them
everywhere.\\
Note that weakly (i.e. on the constraint surface) we have, requiring the
Gauss constraint to hold even at infinity,
\beq f^1-2f^2-\sqrt{2}a_3=0 \; . \eeq
It is convenient first to compute the asymptotic form of the Einstein
magnetic fields
\begin{eqnarray}
B^1 &\rightarrow& -\frac{\sqrt{2}a_3}{x}+O(1/x^2), \nonumber \\
B^2 &\rightarrow& -\frac{a_3}{x^2}+O(1/x^3), \nonumber \\
B^3 &\rightarrow& \frac{a_2+\sqrt{2}a_1}{x^2}+O(1/x^3)
\end{eqnarray}
to conclude for the vector constraint
\beq V\rightarrow-\sqrt{2}\frac{a_2+\sqrt{2}a_1}{x}+O(1/x^2) \; .\eeq
Hence we have to impose
\beq a_2+\sqrt{2}a_1=0 \eeq
in order to make this functional
finite and differentiability can be achieved by adding the ADM-momentum
as in I. Finally, it is easy to see that with this restriction the scalar
constraint functional is already finite and functionally differentiable
when adding the ADM-energy, provided we set the cosmological constant
equal to zero.\\
Now it is possible to make the restriction that comes from requirement 1)
more concrete. We have
\begin{eqnarray}
0 & = & da_1\wedge df^1+\sqrt{2}(da_2\wedge df^2+da_3\wedge df^3)\nonumber\\
  & = & -\frac{1}{\sqrt{2}}da_2\wedge d(f^1-2f^2)+\sqrt{2}da_3\wedge df^3)
      \nonumber\\
  & = & -\frac{1}{\sqrt{2}}da_2\wedge d(f^1-2f^2-\sqrt{2}a_3)
        +da_3\wedge d(\sqrt{2}f^3+a_2) \; .
\end{eqnarray}
Note that the bracket of the 1st wedge product in the last line of (3.16)
vanishes weakly according to (3.12). Hence it is consistent with the
constraint equations to impose
\beq \sqrt{2}f^3+a_2=0\; . \eeq
\\
This completes the boundary conditions at the ends of the hypersurface. What
about the interiour, the compactum K ? For gauge transformations one
requires that the asymptotic structure is untouched. Since the compactum
K is also such a kinematical ingredient of the formalism, we also require
that for gauge transformations the Lagrange multipliers have compact support
{\em outside and inside} the compactum K while for symmetries they shall be
smooth functions on all of $\Sigma$. Hence there is a transition region
between the asymptotic ends and the compactum K. As for the fields, it is
motivated to adapt their behaviour in K in such a way that {\em observables}
are well-defined. We therefore have to postpone this item at this stage and
come back to it after the formal expressions for the observables have been
found.\\
Note that in the literature one usually assumes that 'there exists
a regular initial data set on the hypersurface' (\cite{20}). Since initial
data are in one to one correspondence with the Dirac observables, what we
do here in choosing the boundary conditions in the interiour is nothing
else than a realization of this assumption in a concrete example.\\
Accordingly, the definition of the fall-off behaviour of the fields becomes
(partly) a dynamical ingredient of the formalism.\\
We have by now succeeded to give a definition of the phase space which relies
on minimal requirements and which is general enough to allow for non-trivial
dynamics on the reduced phase space (compare also the 2nd paper in I).\\
\\
b) Compact topologies\\
\\
Here it is sufficient to require the fields and Lagrange multipliers to be
smooth and finite everywhere. The cosmological constant may take any finite
value. Obviously, the case of compact topologies is much more easier to
handle from a technical point of view.

\section{Symplectic reduction of the model}

We will use some basic facts from the theory of symplectic reduction which
can be looked up in the second paper of I and in much greater detail in
\cite{7}, \cite{10}.
\\ We can apply that theory here because, as was shown in I, the present
model is a field theory with first class constraints.\\
According to that theory we are thus first of all interested in the solutions 
of constraint 
equations.\\
\\
Recall from I that the following set of 'cylinder' coordinates was
suggested from the transformation properties of the gravitational variables
under the gravitational Gauss constraint
\beq (A_2,A_3)=\sqrt{A}(\cos(\alpha),\sin(\alpha)),\;(E^2,E^3)=
\sqrt{E}(\cos(\beta),\sin(\beta))\;. \eeq
Now, recall the following result from I :\\
Lemma : The reduced symplectic potential with respect to the Gauss
constraint is given by
\beq i\kappa\Theta[\partial_t]=\int_\Sigma dx(\dot{\gamma}\pi_\gamma
+\dot{B^1}\pi_1+\dot{\omega}p) \; , \eeq
where 
\beq \gamma:=A_1+\alpha',\pi_\gamma:=E^1,B^1:=\frac{1}{2}(A-2)\;
\mbox{and}\; \pi_1:=\sqrt{E/A}\cos(\alpha-\beta). \eeq
Proof : compare I, second paper\\
\\
In the following p will already be taken as a constant. Also we will deal
with an arbitrary cosmological constant for the sake of generality.
We take then the following linear combinations of the vector and the
scalar constraint functional
\begin{eqnarray}
 E^1 E^2 V+E^3 C & = & E(E^1 B^3+\frac{1}{2}E^3
 (\frac{\kappa p^2}{2E^1}+\kappa\lambda E^1+B^1)) \nonumber\\
-E^1 E^3 V+E^2 C & = & E(E^1 B^2+\frac{1}{2}E^2
(\frac{\kappa p^2}{2E^1}+\kappa\lambda E^1+B^1)) \; ,
\end{eqnarray}
where $E=(E^2)^2+(E^3)^2$.\\
Setting these expressions strongly zero we obtain, exactly as in I, 2 possible
solutions :\\
Case I : $E=0$ (degenerate case)\\
Looking at the formula for the metric (3.2) we see that there is no radial
distance now. From the reality of the triads we conclude further that
$E^2=E^3=0$ whence we conclude $E^1=E^1(t)$ via setting the Gauss constraint
equal to zero. Obviously this solution of the constraint equations is not
valid in the asymptotic ends since it violates the asymptotic conditions on
the fields. It can therefore only hold in the compactum K. For compact
topologies it is a global solution of the constraints. We can thus apply
the above framework of symplectic reduction only for that part of the
symplectic potential which corresponds to K or $S^1$ and obtain the reduced
symplectic potential
\beq \hat{\Theta}[\partial_t]=E^1\frac{d}{dt}(\frac{-i}{\kappa}\int_M dx A_1)
+p\frac{d}{dt}\int_M dx\omega=:m\dot{T}+p\dot{\Phi} \eeq
where M means K or $S^1$.\\
Case II : $E\not =0$ (nondegenerate case)\\
We now conclude
\begin{eqnarray}
0 & = & E^1 B^3+\frac{1}{2}E^3(\frac{\kappa p^2}{2E^1}
+\kappa\lambda E^1+B^1), \nonumber\\
0 & = & E^1 B^2+\frac{1}{2}E^2
(\frac{\kappa p^2}{2E^1}+\kappa\lambda E^1+B^1)
\end{eqnarray}
and can further distinguish between a) $f=0$ and b) $f\not=0$ where
$f=\frac{\kappa p^2}{2E^1}+\kappa\lambda E^1+B^1$. We assume the generic
case $p\not=0$ to hold in the following. Then in order that the scalar
constraint makes sense at all, we must have $E^1\not=0$. \\
Subcase a)\\
Here it follows from (4.6) that $B^2=B^3=0$. Hence, from the
Bianchi-identity
$A_3 B^2-A_2 B^3-(B^1)'=0$, we infer $(B^1)'=0$ i.e. $B^1$ is a spatial
constant whence from $f=0$, $E^1$ is also a
spatial constant. Therefore, this solution of the constraint equations can
also only refer to the compactum K or to the compact case. Finally, we have
$0=A_2 B^2+A_3 B^3=A\gamma=2(1+B^1)\gamma\;\Rightarrow\;\gamma=0\mbox{ or }
B^1=-1$. Using the above lemma we
can finally carry out the pull-back on $M\;\in\;\{K,S^1\}$ :
\beq \hat{\Theta}[\partial_t]=\dot{B}^1(\frac{-i}{\kappa}\int_M dx \pi_1)
+p(\frac{d}{dt}\int_M dx \omega)=:m\dot{T}+p\dot{\Phi} \;. \eeq
or
\beq \hat{\Theta}[\partial_t]=\pi_\gamma\frac{d}{dt}(\frac{-i}{\kappa}\int_M
dx\gamma)+p(\frac{d}{dt}\int_M dx \omega)=:m\dot{T}+p\dot{\Phi} \;. \eeq
We do not consider the trivial case $A=\gamma=0$ which is equivalent
to 2-dimensional pure Maxwell theory without dynamics.\\
Subcase b)\\
We can, by virtue of $f\not=0$, divide by f to solve eqs. (4.6) for the
momenta $E^2\;\mbox{and}\;E^3$
\begin{eqnarray}
E^2 & = & -\frac{2 (E^1)^2}{\kappa(p^2/2+\lambda(E^1)^2+B^1 E^1} B^2,
\nonumber\\
E^3 & = & -\frac{2 (E^1)^2}{\kappa(p^2/2+\lambda(E^1)^2+B^1 E^1} B^3
\end{eqnarray}
and insert this into into the Gauss constraint :
\beq 0=(\kappa(p^2/2+\lambda(E^1)^2)+B^1 E^1)(E^1)'+2(E^1)^2(B^1)' \; .\eeq
Eqn. (4.10) can be written as the derivative of a constant function of
$B^1\;\mbox{and}\;E^1$ with respect to x after multiplying it with the
integrating multiplicator $(E^1)^{-3/2}$ :
\begin{eqnarray}
& & (\frac{1}{\sqrt{E^1}}[\kappa(-p^2/2+\lambda(E^1)^2/3)+B^1 E^1])'=0\\
&\mbox{i.e.} & [\kappa(-p^2+\lambda(E^1)^2/3)+B^1 E^1]^2=P E^1 \; .
\end{eqnarray}
The integration constant, P, is real because, as proved in I, the magnetic
fields are (weakly) real.
Note further that the rhs. of (4.15) is non-negative whence $P E^1\ge0$. In
the asymptotical ends $E^1$ is positive such that P is non-negative and thus
can be written $P=m^2\;\mbox{where}\;m$ is real and it is easy to see the
relation of m with the gravitational mass $m_G$ : using (3.4), (3.13) and 
(4.16) and expanding in powers of r one finds $m^2=(-\sqrt{2}a_3)^2$
together with $a_3=-\sqrt{2}m_G$ and we will fix the sign ambiguity by
requiring that $m=+\sqrt{2}a_3$.\\ 
If one uses the positive
censorship conjecture that there are no naked singularities, then the
positive energy theorem (see ref. \cite{11}) tells us that m is not positive 
because, although the energy density of matter $E/(2E^1)p^2$ is manifestly
non-negative, in the positive m case the singularity at the origin $x=0$ is
timelike and therefore no spacelike hypersurface with everywhere regular
initial data exists such that the positive energy theorem does not apply. We
will, however, not make such an assumption (that case is treated in the 2nd
paper of I).\\

Note that (4.12) is a purely algebraic equation of fourth order for $E^1$ in 
terms
of $B^1$. Although algebraic eqs. of fourth order can be solved analytically,
the corresponding formulas are too complicated as that the theorem in I
could be applied (solving the Hamilton-Jacobi equations by quadrature
techniques) in the general case. An exception is the special case $P=p=0$ :
in that case the electric fields are proportional to the magnetic ones
\beq E^I=-\frac{3}{\kappa\lambda}B^I\;, I=1,2,3 \eeq
and the unique solution of the Hamilton-Jacobi equations
$\delta S/\delta A_I=-3/(\kappa\lambda)B^I$ is {\em precisely} the reduction
to spherical symmetry of the SO(3)-Chern-Simons functional
\beq S=-\frac{6}{4\pi\kappa\lambda}\int_{\Sigma^3}d^3x tr[A\wedge(F
-\frac{1}{3}A\wedge A]=-\frac{3}{\kappa\lambda}\int_\Sigma dx\gamma B^1 \eeq
which was to be expected from the corresponding result for the full theory
(see ref. \cite{12}). However, since the solution of the Hamilton-Jacobi
equation does not depend on any free parameter, it follows that the
reduced symplectic potential vanishes in that case. This solution has thus
only the status of a total differential that can be added to the action and
gives rise to a $\theta$-term as in Yang-Mills theory.\\
Note also that formula (4.12) can be obtained by the method of CDJ (see
ref. \cite{13}) restricted to spherical symmetry (see I), however its
derivation is not simplified by that method, so we do not display it here.
That the CDJ-method only applies for case II.b follows from the fact that
only then 'electric and magnetic metric' are non-degenerate.\\

In order to actually reduce the theory in the general case, one can proceed
as\\ follows : the basic observation is that formula (4.12) can be solved 
easily
for the magnetic field $B^1$ which, according to the above lemma, {\em can be
chosen as a canonical coordinate}. Thus, if one simply changes the
polarization so that $B^1$ becomes a momentum, the chance that one can
complete the reduction becomes significantly larger. Accordingly, let us
write formula (4.12) as
\beq B^1=\frac{\kappa p^2}{2\pi_\gamma}-\frac{\lambda\kappa}{3(\pi_\gamma)^2}
+\frac{m}{\sqrt{\pi_\gamma}} \;, \eeq
where we have confined ourselves to an asymptotic region in order to have
P positive or zero, the compact case can be treated analogously.
From the transformation properties of the fields under diffeomorphisms (see
I) we know that the Gauss-reduced vector constraint must be
\beq V=-\gamma\pi_\gamma'+(B^1)'\pi_1 \;\eeq
which can also be checked explicitly, of course. Substituting for $B^1$
from (4.15) we can solve (4.16) for $\gamma$ :
\beq \gamma=\pi_1(-\frac{\kappa p^2}{2(\pi_\gamma)^2})+\frac{2\lambda\kappa}
{3(\pi_\gamma)^3}-\frac{m}{2(\sqrt{\pi_\gamma})^3} \;. \eeq
Since for the asymptotically flat topologies we must have $\lambda=0$ we
conclude from (4.15) that asymptotically $E^1\propto(B^1)^{-2}\propto x^2$,
i.e. we have finally found a solution of the constraints that fits into the
fall-off requirements valid for the asymptotic regions. Furthermore $E^1>0$
asymptotically, so $P:=m^2$ is positive while m is real and thus $E^1\ge0$
in the whole asymptotic region.\\
The last step is then to pull back the symplectic potential. We comprise
total differentials that appear during the reduction process in a functional
S after having displayed them. Then we have
\begin{eqnarray}
(\iota^*\Theta)[\partial_t] & = & -\iota^*[\int_\Sigma dx(\dot{p}\omega
-i/\kappa(\dot{\pi}_\gamma\gamma+\dot{\pi}_1 B^1)) \nonumber\\
& & +\frac{d}{dt}\int_\Sigma dx(p\omega-i/\kappa(\pi_\gamma\gamma+\pi_1 B^1))]
\nonumber \\
& = & -\int_\Sigma dx(\dot{p}\omega-i/\kappa\dot{\pi}_\gamma\pi_1
(-\frac{\kappa p^2}{2(\pi_\gamma)^2}+\frac{2\lambda\kappa}{3(\pi_\gamma)^3}
\nonumber\\
 & &-\frac{m}{2(\sqrt{\pi_\gamma})^3})-i/\kappa\dot{\pi}_1(\frac{\kappa p^2}
{2\pi_\gamma}-\frac{\lambda\kappa}{3(\pi_\gamma)^2}+\frac{m}
{\sqrt{\pi_\gamma}}))+\dot{S}\nonumber \\
& = & -\dot{p}\int_\Sigma dx\omega+i\frac{p^2}{2}\int_\Sigma dx(-
\frac{\dot{\pi}_\gamma\pi_1}{(\pi_\gamma)^2}+\frac{\dot{\pi}_1}{\pi_\gamma})
+im/\kappa \int_\Sigma dx(-\frac{\dot{\pi}_\gamma\pi_1}{2(\sqrt{\pi_\gamma})^3})
\nonumber\\ & & +\frac{\dot{\pi}_1}{\sqrt{\pi_\gamma}})
+i/\kappa\frac{\lambda\kappa}{3}\int_\Sigma dx(\frac{2\dot{\pi}_\gamma\pi_1}
{\pi_\gamma)^3}-\frac{\dot{\pi}_1}{(\pi_\gamma)^2})+\dot{S} \nonumber \\
& = & -\dot{p}\int_\Sigma dx\omega+i\frac{p^2}{2}\frac{d}{dt}
\int_\Sigma dx\frac{\pi_1}{\pi_\gamma}+im/\kappa \frac{d}{dt} \nonumber\\
& & \int_\Sigma
dx\frac{\pi_1}{\sqrt{\pi_\gamma}}-i/\kappa\frac{\lambda\kappa}{3}\frac{d}{dt}
\int_\Sigma dx
\frac{\pi_1}{\pi_\gamma)^2}+\dot{S} \nonumber \\
& = & \dot{p}\int_\Sigma dx(-i\frac{p\pi_1}{\pi_\gamma}-\omega)
+\dot{m}\int_\Sigma dx(-i/\kappa)\frac{\pi_1}{\sqrt{\pi_\gamma}}
-i/\kappa\frac{d}{dt}\int_\Sigma dx(\frac{\lambda\kappa}{3} \nonumber\\
& & \frac{\pi_1}{(\pi_\gamma)^2}-\frac{\kappa p^2\pi_1}{2\pi_\gamma}
-\frac{m \pi_1}{\sqrt{\pi_\gamma}})+\dot{S} \nonumber \\
& =: & \dot{p}\Phi+\dot{m}T+\dot{S}
\end{eqnarray}
where we have assumed that the cosmological constant is time-independent
(otherwise we introduce a new time variable $\tau$ according to
$d\tau(t)/dt=\lambda^{-1}(t)$ and absorb a factor of $1/\lambda$ into the
variables $T\;\mbox{and}\;\Phi$ which is, of course, only possible if
$\lambda(t)$ is nowhere vanishing).

\section{The classical *-algebra}

\subsection{Finiteness and functional differentiability}

Let us first check whether the formal integral expressions for
$T\;\mbox{and}\;\Phi$
derived in the last section, which serve as candidates for the coordinates of
the reduced phase space, are actually finite and functionally differentiable
with respect to the coordinates of the large phase space.
\\
Case I :\\
Finiteness is trivial since the integrals involved are over a compact set
and we require the fields $A_1\;\mbox{and}\;\omega$ to be smooth and finite
there. Functional differentiability is also trivial since no spatial
derivatives appear in the integrands of $T\;\mbox{and}\;\Phi$.\\
Case II.a :\\
$\gamma=0$ : since we have $A=2(B^1+1)=
\mbox{const.}\not=0$ in general, the factor $\sqrt{E/A}$ is smooth and finite
in general. Furthermore, it follows from the Gauss constraint
\beq 0=(E^1)'=\sin(\alpha-\beta)\sqrt{AE}, \eeq hence $\cos(\alpha-\beta)=1$.
Since we again integrate over a compactum, T is again finite in general if
we impose the fields A and E to be finite and smooth there. Since no spatial
derivatives are involved, functional differentiability is trivial.\\
$A=0$ : we require $\gamma$ to be smooth and finite, so finiteness of T
is no problem because we integrate it over a compactum. For T to be
functionally differentiable, we need it to vanish at the boundaries of K.\\
Case II.b :\\
First of all, let us express the observables m and p as functionals on the
phase space. To that end, let D be a scalar density of weight one with
respect to the natural metric on $\Sigma$ (derived from its distance
function, available because $\Sigma$ is a normed space) normalized to 1, i.e.
\beq \int_M dx D(x)=1\; \eeq
(one could also choose D to depend on the fields, e.g. $D=(A/2)'$).
Then one can express m and p as
\begin{eqnarray}
m & := &\int_M dx D \frac{B^1 E^1-\kappa(p^2/2-\lambda/3(E^1)^2)}
{\sqrt{E^1}},\nonumber \\
p & = & \int_M dx D p .
\end{eqnarray}
These functionals are obviously finite and functionally differentiable on
the whole phase space because D is at least $O(1/x^2)$, the rest of the
integrands is $O(1)$ and there are no spatial derivatives involved.\\
Because
of the subtlety related to the support of the Lagrange multipliers, m
and p may take different constant values in the disconnected parts of the
support of the Lagrange multipliers on $\Sigma$ corresponding to gauge
transformations. Between these regions of constancy, in the transition
region between K and the asymptotic ends, m and p should change smoothly.
Now it would be satisfactory if these independent, different possible values
of m in the different regions of $\Sigma$ would correspond to {\em different
canonical variables} because the various asymptotic regions should
correspond to different asymptotic observers. This is a physical
motivation and no mathematical prediction of the formalism ! Readers that
feel uneasy may skip the following paragraph in which we
introduce some further structure alluded to at the end of section 3 in order
to achieve this aim, and may view m and p as constant over all of $\Sigma$
in the sequel.\\
We will simply require in addition to the restrictions mentioned in section
3 the support of the integrands of T and $\Phi$ to avoid the transition
region and the origin in K (the same for the Lagrange multipliers, even for
symmetries). The part of $\Sigma$ which is neither origin
nor a transition region will be referred to as {\em support region}.
Since there are 8 basic fields but only 4 reduced coordinates
per region, the smoothness and support requirements on the integrands
of the latter can easily be satisfied.

Then the reduced symplectic potential splits into a sum over
ends and the compactum
\beq \hat{\Theta}=\dot{p}_K\Phi_K+\dot{m}_K T_K+\sum_{A=1}^n(\dot{p}_A\Phi_A
+\dot{m}_A T_A) \eeq
because on the support of the integrands $i_T\mbox{ and }i_\Phi$ of T and
$\Phi$, m and p are constrained to be spatially constant and can therefore be
dragged out of the integral. This furnishes our aim.
Due to our support conditions, the
degrees of freedom of the fields in the transition regions are so to say
frozen : the support conditions are never changed under evolution because
the right hand side of the equations of motion (6.1) vanish trivially if x
happens to fall into the transition region since by definition the Lagrange
multipliers do not have their support there, so even {\em all} the field
variables do never alter their value there. Of course, this means that our
family of hypersurfaces actually does not form a foliation because those parts 
of $\Sigma$ where the lapse has no support will never move. This, eventually,
will also restrict the spacetime manifold if the hypersurfaces are to stay 
spacelike.\\
Note also that because of the just
introduced support conditions on the integrands of T and $\Phi$ there
will never arise boundary terms corresponding to K when varying T and $\Phi$.
Finiteness and functional differentiability are then already assured for
the the variables corresponding to K.\\
All calculations in the rest of this paragraph will therefore deal with
respect to one
asymptotic end only in order to avoid the labelling of the various ends i.e.
we set $M:=\Sigma_A$ for some $A\;\in\;1..n$ in the sequel.\\
Since $\omega=O(1/x^2)$ we can focus at the other expressions in the
asymptotic regions and we have $\lambda=0$ (in the case of $\Sigma=S^1$ we
are already done). Hence we need to determine the fall-off behaviour of
$\pi_1$. Let
\beq g:=-2 (E^1)^2/(\kappa[p^2/2+\lambda(E^1)^2]+B^1 E^1), \eeq then
we have $E^2=g B^2,\;E^3=g B^3,\;\mbox{and}\;(E^1)'=g(B^1)'$ on the
constraint surface. Computing $E$ we are interested in
\beq (B^2)^2+(B^3)^2=\frac{1}{A}([(B^1)']^2+(A\gamma)^2), \eeq
a formula which was obtained by writing the magnetic fields in terms of
'cylinder coordinates'. On the other hand we have
\begin{eqnarray}
& & \alpha-\beta=\arctan(A_3/A_2)-\arctan(B^3/B^2) \nonumber\\
& = &\arctan(\frac{A_3 B^2-A_2 B^3}{A_2 B^2+A_3 B^3})
=\arctan(\frac{(B^1)'}{A\gamma}).
\end{eqnarray}
We now express the cosine function in terms of the tangens function,\newline
$\cos(x)=\pm 1/\sqrt{1+\tan^2(x)}$, to conclude
\beq \cos(\alpha-\beta)=\pm\frac{A\gamma}{\sqrt{((B^1)')^2+(A\gamma)^2}}.\eeq
So we end up finally with
\beq \pi_1=\sqrt{\frac{E}{A}}\cos(\alpha-\beta)=\pm g\gamma \;.\eeq
However, only the upper sign is appropriate since also $\pi_1=
(A_2 E^2+A_3 E^3)=g\gamma$.\\
Looking at the expression for g in terms of $B^1\;\mbox{and}\;E^1$ we find
that $g=O(x^3)$ for $\lambda=0$. Furthermore
\beq \gamma=\frac{A_2 B^2+A_3 B^3}{2(1+B^1)}=O(1/x^3) \eeq
when recalling formula (3.13). Thus $\pi_1=O(1)$ and the integral involving
$\pi_1$ in the expression for $\Phi$ is already finite due to $E^1=
\pi_\gamma=O(x^2)$ whereas the integral T seems to be logarithmically
divergent. However, it is possible to prove that the divergent part
vanishes on the constraint surface. This is not unexpected since an
observable on the full phase space is in any case only determined up to the
addition of a linear combination of the constraint functionals. In order to
obtain a manifestly finite
expression for T on the full phase space we have therefore to add a term
which is proportional to the constraints and which also diverges off the
constraint surface.\\
A  first hint how this expression should look like gives the observation that
if one replaced $A_3\;\mbox{by}\;A_3-\sqrt{2}$ in $A\gamma=A_2 B^2+A_3 B^3$
then the part of the integrand of T $\propto\;(A_3-\sqrt{2})B^3$ would
already be $O(1/x^2)$ and therefore finite. Accordingly, it is motivated
to look for a linear combination of constraints that precisely accomplishes
for that subtraction of $\sqrt{2}$ from $A_3$. The idea is thus to subtract
from T the expression
\beq (-i/\kappa)\int_\Sigma dx \frac{g\sqrt{2}}{\sqrt{E^1}A}(B^3-g^{-1}
E^3) \eeq
since the bracket term is constrained to vanish. Recalling from section 3
how this bracket term could be obtained in terms of the constraints,
formula (4.4), we propose the improved expression for T
\begin{eqnarray}
T_{finite} & := & T+ i/\kappa\int_M dx \frac{g\sqrt{2}}{(\sqrt{E^1})^3 E A}
                       (E^1 E^2 V+E^3 C) \nonumber\\
 & = & (-i/\kappa)\int_M dx \frac{g}{\sqrt{E^1}A}[(A_3-\sqrt{2})B^3
 +(A_2 B^2-\sqrt{2}g^{-1} E^3)] \; .
\end{eqnarray}
The prefactor of the square bracket in the last line of (5.12) is $O(x^2)$,
the first term in the bracket is $O(1/x^4)$ by construction and finally we
have for the leading order part of the second term in the bracket
\begin{eqnarray}
& & A_2 B^2-g^{-1}\sqrt{2} E^3\rightarrow(-\frac{a_2 a_3}{x^3}+O(1/x^4))-
(\frac{\sqrt{2}a_3 f^3}{x^3}+O(x^4))\nonumber\\
& = & -\frac{a_3(a_2+\sqrt{2}f^3)}{x^3}+O(1/x^4)=O(1/x^4)
\end{eqnarray}
i.e. $T_{finite}$ is already a {\em finite} functional on $\Gamma$ due to
the important eqn. (3.17). It is interesting to see that the finiteness
of the symplectic structure enforces the finiteness of the observable
$T_{finite}$ and how fine tuned this finiteness comes about to hold !\\
Next we come to discuss functional differentiability : \\
The only terms that could spoil differentiability are those that appear
with spatial derivatives in the integrand of $T_{finite}\mbox{ or }\Phi$ 
because they give
rise to a boundary term in a variation of these functionals. Let us study
these boundary terms (note that \\
$g\rightarrow(x^3\sqrt{2})/a_3+O(x^2)$) :
\begin{eqnarray}
& & i\kappa(\delta\Phi)_{|boundary\;term}= \kappa p\int_{\partial M}
\frac{g}{E^1 A}(A_2\delta A_3-A_3\delta A_2)=-\frac{p}{a_3}
\int_{\partial M}\delta a_2\nonumber \\
& & i\kappa(\delta T_{finite})_{|boundary\;term}=\int_{\partial M}
\frac{g}{\sqrt{E^1}A}(A_2 \delta A_3-(A_3-\sqrt{2})\delta A_2)\nonumber\\
& &=\int_{\partial M}\frac{1}{\sqrt{2}a_3}(a_2\delta a_3-a_3\delta a_2)\;.
\end{eqnarray}
Hence both observables fail to be functionally differentiable. Even more
serious : the boundary term of the variation is not exact and it seems that
one is not able to add a counterterm in order to restore differentiability.
What saves the day is that it follows 
from the transformation law of the fields to require $\delta a_2=
\delta a_3=0$ which has the consequence that $\Phi\mbox{ and }T_{finite}$
already become differentiable :\\ 
looking at the variation of $A_2,A_3$ as derived from computing the
Poisson brackets with the constraint functionals (which are already
functionally differentiable by construction of the phase space in section
3.2), equation (6.1), and inserting the asymptotic behaviour of the
various fields, we conclude $\delta a_2=\delta a_3=0$ even for the
symmetry transformations that we allow.
The result $\delta a_2=\delta a_3=0$ means that the dynamical part (due to
$\sqrt{2}a_1+a_2=0$) of $A_I$ rests in the higher order terms in their
asymptotic expansions.

\subsection{Reality conditions}

In all cases the set of coordinates $E^I,\omega,p$ is real. Therefore, in
cases I and II.a, $\Phi$ is already known to be real.
The range of $E^I$ is not always the whole real axis (see below).\\
\\
Case I :\\
The reality condition for $A_1$ is given by $\bar{A}_1+A_1=2\Gamma_1\;
\mbox{where}\;\Gamma_1$ is given by $-\beta'$ (see I). Obviously, there is a
problem because $\beta=\arctan(E^3/E^2)\;\mbox{but}\;E^2=E^3=0$. Let for
example $E^1=h+k,E^2=h e^2,E^3=h e^3$ where $e^2,e^3,h$ are arbitrary real
smooth functions and k is a spatial constant. Then the constraint surface is
defined by $h=0$ however $\beta$
is ill-defined. This shows that Ashtekar's formalism is {\em} not, although
frequently said so (ref. \cite{4}), really an extension of Einstein's theory
in the sense that it allows for degenerate metrics, because the degeneracy
makes the reality conditions ill-defined (this is also true for the full
theory because the spin connection is a homogeneous function of degree zero
in terms of the densitized triads). In order to have a real reduced theory
we may motivate to set $\beta=const.$ so that T becomes real. In that case
we would have a cotangent bundle over $R^2$ as the reduced phase space for the
compactum K. There is no reality condition on $A_2,A_3$ because we have
no motivation to give the spin-connections $\Gamma_2,\Gamma_3$ a definite
real value.\\
\\
Case II.a :\\
The solution of $f=0$ is given by
\beq E^1=-B^1/(2\kappa\lambda)\pm\sqrt{(B^1/(2\kappa\lambda))^2-
p^2/(2\lambda)}\;\mbox{for}\;\lambda\not=0 \eeq whereas for $\lambda=0$
we have $E^1=-p^2/(2B^1)$. Hence the range of $E^1$ depends here on the value
of the cosmological constant and there is also the constraint ($B^1\le-1
\mbox{ for } \gamma=0$, see below)
\beq B^1\le \mbox{min}(\{-1,-\sqrt{2\lambda}|p|\kappa\})\;\mbox{for}\;
\lambda>0\;
\mbox{(and only}\;B^1\le -1\;\mbox{for}\;\lambda<0)
\eeq in order to guarantee the reality of $E^1$. Hence, since in the case
$B^1=-1$ of subcase II.a the reality of $E^1$ cannot be guaranteed for 
$\lambda\not=0$ we 
we have to stick with $\gamma=0$ in that case. For $\lambda=0$ both cases
$\gamma=0\mbox{ and }B^1=-1$ are possible. \\
More precisely the ranges of $E^1, B^1,p$ are linked as follows for the 
various values of the cosmological constant :\\ 
i)   $\lambda<0$ : $E^1$ is monotonously decreasing (increasing) with
decreasing $B^1$ for the lower (upper) sign and therefore its upper (lower)
bound, which is negative (positive), is given by inserting the
upper bound of $B^1$ into formula (5.18). There is thus a gap in the range
of $E^1$ of width $2\sqrt{(B^1/(2\kappa\lambda))^2-p^2/(2\lambda)}$
symmetrically around zero.\\
ii)  $\lambda=0$ : $E^1$ is positive but bounded from above by $p^2/2$.\\
ii) $\lambda>0$ : $E^1$ is monotonously increasing with decreasing $B^1$
and therefore its lower bound, which is positive, is given by inserting the
upper bound of $B^1$ into formula (5.18) with the negative sign.\\
Now let us discuss the two subcases of II.a :\\
$\gamma=0$ :\\
Since $E\not=0\;\mbox{but}\;(E^1)'=0$ it follows that $\Gamma_2:=-(E^1)'E^3/E
=\Gamma_3:=(E^1)'E^2/E=0$ whence $A_2,A_3$ are imaginary. Thus, $\sqrt{E/A}=
\pi_1$ is imaginary, i.e. T is again real while $m=B^1=A/2-1$ is bounded from
above by -1 for $\lambda\le 0$ and by $\mbox{min}(-1,-\sqrt{2\lambda}|p|
\kappa)$ for
$\lambda>0$. 
Thus, we obtain the cotangent bundle over the half-plane $R\times R^{\le-1}$
for $\lambda=0$ or over the 'cut wedge' $\{(p,B^1);\;p\;\in\;R,\;B^1\le\;
\mbox{min}(-1,
-\sqrt{2\lambda}\kappa|p|\}$ as the reduced phase space respectively.\\
$B^1=-1$ : \\
Since $A=0$ we have $A_2=\pm i A_3\mbox{ whence }\alpha'=0
\mbox{ i.e. }\gamma=A_1$. Furthermore $B^3=\mp i B^2=(A_3)'\pm i A_1 A_3=0
\mbox{ whence }A_1=\pm i[\ln(A_3)]'$ i.e. $\gamma$ is imaginary because
$A_3$ is imaginary and thus we conclude $\Gamma_1=-\beta'=0$. Hence
T is real with range over the whole real axis while $m=E^1<p^2/2$ because
we have $\lambda=0$.\\
The reduced phase space is therefore the cotangent bundle over 
$\{(p,m); \; p\in R,\;m<p^2/2\}$.\\
\\
Case II.b :\\
In I it was proved that the function $\gamma$ is (weakly) imaginary, hence
$\pi_1$ is imaginary while $\pi_\gamma=E^1$ is real. Accordingly, the
integrands of T and $\Phi$ are both real.\\
The reduced  phase space can thus be described as follows : in every
asymptotic end A we have a cotangent bundle over $R^2$ as well as in K.\\
\\
Note that we also could glue together case II.b in the asymptotic regions
with one of the cases I and II.a in the compactum for open topologies
while we have to make a choice for compact topologies between the 3 cases.
Of course, one could also imagine to have arbitrarily many regions in
which one of the 3 cases holds (for both types of topologies) but in order
to make the associated observables again differentiable one would have
to impose additional structure (support conditions for these additional
regions) which we do find unnatural.

\subsection{Geometry of the constraint surface}

We have obtained, over each of the regions of $\Sigma$, up to three
apparently unrelated constraint surfaces and reduced phase spaces.\\
The question arises whether various observables on the 3 different constraint
surfaces should be treated as independent of each other or not. This is an
important question because, as already pointed out in I, it affects the
dimension of the reduced phase space.\\
From the way we found the constraint surface it is clear how this split
into apparently three different leaves came about : by taking linear
combinations of the original constraint generators we find first equivalent
constraint generators but in such a way that each of them
can be written as a product. A product vanishes if any of the factors
vanishes. Hence we obtain new inequivalent constraint generators depending
on which set of factors we chose to vanish. These new sets of constraints
form again, as one can show,  a 1st class algebra and the symplectic reduction 
works for every
set separately. However, the flow of the constraint generator corresponding
to one of these sets lies only tangential to the constraint surface defined
by this set and thus never penetrates the other parts of the complete
constraint surface (corresponding to the other set) except for possible
intersection subsets between the different leaves of the constraint surface.
Now, points of the full constraint surface which lie on the same flow line
of a Hamiltonian vector field corresponding a constraint generator are
to be identified. Accordingly, intersection points will lead to an
identification of some points in the two reduced phase spaces.\\
Once one has obtained a (partial) identification of points of the various
leaves of the reduced
phase space, one can glue these together along these
points and one obtains one big reduced phase space which consists of leaves
which communicate through the identification process.\\
Altogether, one gets a topology of the complete reduced phase space which
is similar to a Riemann surface with various leaves and
cuts in it.\\
A detailed analysis of that problem, including the general theory of how 
to treat a 
factorizing constraint, is given in \cite{24}, but we will not need these
results here and just restrict ourselves to a short description :\\
One would need to compute the intersection domain
between the leaves of the constraint surface for our model.
It is possible to show that all leaves have non-empty mutual
intersection, that the intersection domain is only presymplectic and that
the resulting reduced phase space has a quite complicated topology.\\
\\
How does one do quantum theory on these glued leaves ? There are two obvious
strategies available :\\
1) Probably the only
constructive way to deal with the problem is to 'hide' the complicated
topology of the reduced phase in a set of relations between the variables
of an overcomplete set and to proceed along the lines of the algebraic
quantization programme (compare third reference of \cite{4} and references
therein).\\
2) One excludes the intersection domain by hand (this is motivated by its
typically presymplectic nature anyway, in particular this is true for our 
model). This disconnects the two reduced
phase spaces by brute force and they can be treated just as phase spaces
of two unrelated theories, that is, separately.\\
\\
We will in quantum theory choose the second strategy.

\section{Proof of conjugacy and gauge invariance and derivation of evolution}

In the derivation of the observables in section 5 we neglected several
boundary terms and total differentials. It is therefore not unnecessary
to check whether the improved quantities of section 6 are really conjugate
variables, if they are really gauge invariant and what their evolution
equations are. Of course, in the case of compact topology without boundary
the following analysis is unnecessary.\\
By construction the symmetry generators are functionally differentiable.
The variation of the coordinates of the full phase space is given by
($G:=G[\Lambda,L,M,U]:=\int_\Sigma dx[(\Lambda-L A_1){\cal G}+L V+M C+U
\;^M{\cal G}]-\int_{\partial\Sigma}[(\Lambda-L A_1)E^1+L((A_3-\sqrt{2})
E^3+A_2 E^2)+M((A_3-\sqrt{2})E^2-A_2 E^3)E^1+Up$ is the full symmetry
generator and we absorb the small variation parameter $\epsilon$ into
the Lagrange multipliers $\Lambda,L,M,U$)
\begin{eqnarray}
\frac{1}{i\kappa}\delta A_1 & := &\frac{1}{i\kappa}\{A_1,G\}=-\Lambda'
+(LA_1)'+M(B^2 E^2+B^3 E^3+E\kappa/2(-p^2/(2(E^1)^2)+\lambda))\nonumber\\
\frac{1}{i\kappa}\delta A_2 & := &\frac{1}{i\kappa}\{A_2,G\}=-\Lambda A_3
+LA_2'+M(E^1 B^2+E^2(B^1+\kappa(p^2/(2E^1)+\lambda E^1))\nonumber \\
\frac{1}{i\kappa}\delta A_3 & := &\frac{1}{i\kappa}\{A_3,G\}=\Lambda A_2
+LA_3'+M(E^1 B^3+E^3(B^1+\kappa(p^2/(2E^1)+\lambda E^1))\nonumber \\
\frac{1}{i\kappa}\delta E^1 & := &\frac{1}{i\kappa}\{E^1,G\}=L (E^1)'
-ME^1(A_2 E^2+A_3 E^3)\nonumber \\
\frac{1}{i\kappa}\delta E^2 & := &\frac{1}{i\kappa}\{E^2,G\}=-\Lambda E^3
+(LE^2)'-(ME^3 E^1)'-M(E^2 E^1 A_1+1/2 A_2 E) \nonumber \\
\frac{1}{i\kappa}\delta E^3 & := &\frac{1}{i\kappa}\{E^3,G\}=\Lambda E^2
+(LE^3)'+(ME^2 E^1)'-M(E^3 E^1 A_1+1/2 A_3 E) \nonumber \\
\delta \omega &:= & \{\omega,G\}=-U'+M E/(2E^1) p\nonumber \\
\delta p & := & \{p,G\}=0 \; .
\end{eqnarray}
The symmetry algebra is given by (on $\bar{\Gamma}$, compare I)
\begin{eqnarray}
& &\frac{1}{i\kappa}\{G[\Lambda_1,L_1,M_1,U_1],G[\Lambda_2,L_2,M_2,U_2]\}_{\bar{\Gamma}}
\nonumber\\
 & = & \int_{\partial\Sigma}[(\Lambda_1 L_2-\Lambda_2 L_1)(A_2 E^3-A_3 E^2)
\nonumber \\
 & & +(\Lambda_1 M_2-\Lambda_2 M_1)(A_2 E^2+A_3 E^3)E^1-(L_1 M_2-L_2 M_1)
\nonumber\\ & &
(A_1 E^1(A_2 E^2+A_3 E^3)+\frac{1}{2}E(B^1+\kappa(p^2/(2 E^1)+\lambda E^1))]
\end{eqnarray}
i.e. it is abelian (recall the boundary conditions to show that the rhs of
(6.2) vanishes identically)
as it should be since we are dealing
only with the translation subgroup of the Poincare group at spatial infinity.
\\
We can now proceed to vary the expressions for the observables. By
construction they are functionally differentiable so we do not need to worry
about boundary terms, the Poisson bracket with G {\em is} a volume integral
again. We are interested in the restriction of the Poisson bracket to the
constraint surface only, hence when varying the part of T which is
proportional to the constraint generators, we do not need to care about
the variations of the prefactors of these generators. Rather, they can be
treated as multipliers so that we can apply formula (6.1) when computing
the action of the symmetry generators on the observables. One might object
that the part of the observables which is proportional to a constraint
generator is by itself a divergent expression so that it is doubtful whether
formula (6.2) can be applied, however, since we compute the variation of 2
divergent expressions whose associated divergent parts cancel each other, our
argument is indeed accurate (recall that we {\em first} integrate over a
finite range of x and {\em then} take the limit).\\
Then we obtain on the constraint surface
\begin{eqnarray}
\delta m & = & \int_\Sigma dx D [ \frac{B^1 E^1+\kappa(p^2/2+\lambda(E^1)^2}
{2(\sqrt{E^1})^3}\delta E^1+\sqrt{E^1}(A_2\delta A_2+A_3\delta A_3)+
\frac{\kappa p}{2\sqrt{E^1}^3}\delta p]\nonumber \\
& = & \int_\Sigma dx D\sqrt{E^1}[-g^{-1}\delta E^1+A_2\delta A_2+A_3
\delta A_3+\frac{\kappa p}{2(E^1)^2}\delta p]\nonumber \\
\delta p & = & \int_\Sigma dx D \delta p \nonumber \\
\delta T & = &(-i/\kappa)\int_{\Sigma}dx[\frac{g}{\sqrt{E^1}}\delta A_1
+(\frac{g^2\gamma}{2\sqrt{E^1}^3}A_2+(\frac{g}{\sqrt{E^1}})'\frac{A_3}{A})
\delta A_2 \nonumber \\
 & & +(\frac{g^2\gamma}{2\sqrt{E^1}^3}A_3-(\frac{g}{\sqrt{E^1}})'
 \frac{A_2}{A})\delta A_3
+\frac{\gamma}{\sqrt{E^1}^3}(3/2 g+\frac{g^2}{2 E^1}(B^1+2/3\lambda
\kappa E^1))\delta E^1] \nonumber \\
 & & + i/\kappa \delta G[\Lambda,N,M,0] \nonumber \\
\delta \Phi&=&(-i/\kappa)p\int_{\Sigma}dx[\frac{\gamma g}{p E^1}\delta p
-\frac{1}{p}\delta\omega+\frac{g}{E^1}\delta A_1
+(\frac{g^2\gamma}{2(E^1)^2}A_2+(\frac{g}{E^1})'\frac{A_3}{A})
\delta A_2 \nonumber \\
 & & +(\frac{g^2\gamma}{2(E^1)^2}A_3-(\frac{g}{E^1})'\frac{A_2}{A})
\delta A_3
+\frac{\gamma}{(E^1)^2}(g+\frac{g^2}{2 E^1}(B^1+2/3\lambda
\kappa E^1))\delta E^1] \; ,
\end{eqnarray}
where in the last line of the equation for $\delta T\;,\;\delta G$ has to be
replaced by $\int_\Sigma dx[\{A_I,G\}\delta E^I-\{E^I,G\}\delta A_I
+\{\omega,G\}\delta p-\{p,G\}\delta\omega$ according to eqn. (6.1) with
the\\ 'Lagrange-multipliers'
\begin{eqnarray}
\Lambda & = & A_1 L\;\mbox{and}\; L=\frac{g\sqrt{2} E^1 E^2}{(\sqrt{E^1})^3
E A}, \nonumber\\
M & = & \frac{g\sqrt{2}E^3}{(\sqrt{E^1})^3 E A}\; .
\end{eqnarray}
We have now all necessary
formulas to compute the Poisson brackets between physically relevant
quantities. The actual computation is rather tedious and the reader is
referred to the appendix. However, these computations again show the
{\em fine tuning} and interrelation between the well-definedness of the
various objects that one is dealing with in general relativity.\\
Computing Poisson brackets among the observables and between observables
and symmetry generators reaffirms the canonical structure that has been
formally derived in the last section and that the observables are really
gauge invariant when choosing the Lagrange-multipliers of compact support.
For symmetry transformations on the other hand we obtain
\begin{eqnarray}
\{m_A,G[\Lambda,N^x,\tiN,U]\} & = & 0, \nonumber \\
\{T_A,G[\Lambda,N^x,\tiN,U]\} & = & N_A, \nonumber \\
\{p_A,G[\Lambda,N^x,\tiN,U]\} & = & 0, \nonumber \\
\{\Phi_A,G[\Lambda,N^x,\tiN,U]\} & = & U_A
\end{eqnarray}
while $m_K,T_K,p_K\;\mbox{and}\;\Phi_K$ are all constant and we have defined
\beq N_A(t):=N(x=\partial\Sigma_A,t)\;,\;U_A(t):=U(x=\partial\Sigma_A,t) \eeq
where $N:=\det(q)^{1/2}\tiN$ is the lapse function. Hence the observables
are invariant under radial translations and O(2)-rotations at spatial
infinity while they react nontrivially under time-translations and phase
transformations at spatial infinity.\\
It is expected but for field theories
not completely obvious that the equations of motion (6.5) coincide with the
equations of motion that follow from the reduced Hamiltonian
\beq H_{red}[m,T,p,\Phi]:=G[\Lambda,N^x,M,U]_{|\bar{\Gamma}}=
\sum_{A=1}^n m_A N_A+p_A U_A \;  \eeq
(for systems with a finite number of degrees of freedom and Hamiltonian H
it is easy to prove that for any functional O on the full phase space holds
\[ \{O,H\}_{|\bar{\Gamma}}=\{O_{\bar{\Gamma}},H_{red}\} \]
if and only if O is an observable) provided that we neglect the O(2) charge
$\int_{\partial\Sigma}\Lambda E^1$ which does not spoil the differentiability
of the Einstein-Gauss constraint for symmetries because the variation of the
charge vanishes anyway. We will do this in the sequel.\\
Note that if we had not made the observables manifestly finite and
functionally differentiable but had computed the various brackets in a naive
way, not caring about boundary terms occurring in variations, then we would
not have obtained the contributions from the constraint part and the
evolution equations would change {\em significantly}. This shows how subtle
the treatment of the asymptotically flat case is (compare the appendix to see
this technically).\\
\\
It is easy to solve the equations of motion (6.5) : introduce functions
$\tau_A(t)\;\mbox{and}\;\phi_A(t)$ defined by
\beq \frac{d}{dt}\tau_A=N_A\;\mbox{and}\;\frac{d}{dt}\phi_A=U_A \; \eeq
Then the solution can be written
\begin{eqnarray}
m_A(t) & = & \mbox{const.}, \nonumber\\
T_A(t) & = & \mbox{const.}+\tau_A(t), \nonumber\\
p_A(t) & = & \mbox{const.}, \nonumber\\
\Phi_A(t) & = & \mbox{const.}+\phi_A(t) \; A=1..n
\end{eqnarray}
i.e. the reduced system adopts the form of an integrable system whereby the
role of the action variables is played by the masses and the charges whereas
their conjugate variables take the role of the angle variables.\\
What now is the interpretation of this second set of conjugate variables 
(compare also the second reference in \cite{1}) ?\\
The interpretation of m and p follows simply from the fact that they can
be derived from the reduced Hamiltonian, i.e. they are the well-known surface
integrals ADM-energy and Maxwell-charge. However, their conjugate partners
are genuine volume integrals and we are not able to write them as known
surface integrals. Nevertheless it is possible to give an interpretation :
Recall that for vanishing shift vector, the $g_{tt}$ component of the
spacetime metric is just given by $g_{tt}=-N^2$ and that one defines the
eigentime of a local observer by $d\tau=\pm\sqrt{-g_{tt}}dt$ where the
upper (lower) sign is valid when dt is future (past) directed (note that
dt is proportional to the normal to $\Sigma$ which is assumed to be
future directed. Hence, for vanishing shift, $d\tau=\pm N dt$, whence it
follows from the solution (6.9) that {\em 'on shell'} the variable $T_A$
is nothing else than the eigentime of an asymptotic observer at spatial infinity of the
end $\Sigma_A$. That T must be a time variable follows also from a
dimensional analysis since m and T are conjugate and m has the dimension of
an energy. The eigentime of an observer is intuitively something
'observable' such that this interpretation sounds quite satisfactory.\\
Next recall that $U=\omega_t$ is the t-component of the Maxwell connection,
i.e. the scalar potential of Minkowski space at infinity, i.e. the vacuum
potential. Now the equation of motion
$-d/dt(-\Phi)=U$ looks like the induction law of electrodynamics ! Hence,
it seems that $-\Phi$ (as it should by a dimensional analysis) should
represent something. like a magnetic flux. Can this formal consideration be given
a physical meaning ? Only approximately : Looking at the part
\beq +\int_{\Sigma_A} dx \omega \eeq of $\Phi$ and comparing it with the
formula for a magnetic Maxwell flux through a surface S, $\int_S
d\wedge\omega=\int_{\partial S} \omega$ we see that our interpretation is
formally correct although the line integral over $\Sigma_A$ is not closed and
it is not possible to find a closed line in $\Sigma^3$ including $\Sigma_A$
such that the integral of $\omega_a dx^a$ reduces to our expression. Note
however, that the expression (6.10) is also the variable conjugate to the
electric field in (1+1) Maxwell theory on a Minkowski background and that
the remainder of $\Phi$ vanishes for Minkowski space (Minkowski space
corresponds to vanishing connection $A_1,A_2,A_3-\sqrt{2}\rightarrow 0,\; E^1
\rightarrow x^2$
i.e. $\gamma=0,g/E^1=-4x^2/(\kappa p^2)$). So, $-\Phi$ can also be interpreted 
as the curved analogue of this observable because the spherically symmetric
reduction leads us to an effectively 2-dimensional spacetime. However, the
dynamics of $\Phi$ is completely different from (1+1) Maxwell theory :
there $\dot{\Phi}\propto p$ whereas here there is no dependence of
$\dot{\Phi}$ on the charge ! \\
This finishes the issue of interpretation of the theory at least for
asymptotically flat topologies.

In the compact case, we have no Hamiltonian
and the observables found are constants of motion. They are, nevertheless,
nonvanishing in general. How can we interpret the theory in that case ?
Here one can apply the theory of deparametrization (see ref. \cite{5}).
In the terminology of that paper, $m,T,p\;\mbox{and}\;\Phi$ are
'time-independent' Dirac observables and we have applied the so-called
"frozen formalism" so far.\\
The application of that theory is beyond the 
scope of the present paper.\\
\\
Let us now mention two objections that were raised in discussions about
the results in the first paper of I and which were resolved in the second
paper of I \\(compare also \cite{24}) :\\
1) For a Reissner Nordstr{\o}m foliation, the observable T indeed vanishes.
This foliation follows from setting the shift equal to zero everywhere
(so-called static foliation) and by choosing $E^1=x^2$ since if we check
whether the gauge choice $E^1=r^2$ is preserved under evolution we find on
the constraint surface (recall formula (6.1))
\beq \delta E^1=\delta x^2\stackrel{!}{=}0=-4x^6 M(1+B^1)\frac{\gamma}
{i\kappa[x^2 B^1+\kappa(p^2/2+\lambda x^4)]} \eeq
from which follows $\gamma=0$ ($B^1$ cannot vanish due to the constraint
(4.18)) i.e. $\pi_1=0$ and hence T=0. This means that nonvanishing T
indicates a deviation from the usual Reissner Nordstr{\o}m-foliation. Now
there seems to appear a problem : according to the equations of motion
(6.5) a vanishing T is not stable under evolution ! However, in the
literature one chooses a static foliation at each instant of time (e.g.
ref. \cite{15}) while here this seems to be dynamically impossible.
The contradiction is resolved in the same way as in the second paper of I,
namely by choosing the lapses at the ends of $\Sigma$ appropriately.\\
\\
2) The (extended) Birkhoff theorem (ref. \cite{15}) states that the
4-diffeomorphism inequivalent solutions of Einstein-Maxwell-theory
reduced to spherical symmetry are only labelled by mass and charge. The space
of solutions of gauge-inequivalent solutions of a field theory on the other
hand are in 1-1 correspondence with the reduced phase space. We, however,
find besides the mass and the charge the eigentime and the flux as additional
observables.\\
The apparent contradiction can be concisely resolved as follows (compare 
\cite{16} for a related phenomenon for Bianchi cosmologies) :\\ 
obviously, following Birkhoff's theorem, the observables T and
$\Phi$ are considered as pure gauge, i.e.
they can be set to zero for all times (looking at the proof given by
Birkhoff, one finds that it is purely geometrical in nature, that is, it does
not care about fall-off properties of diffeomorphisms and fields etc.). From the
Hamiltonian point of view, this is impossible because we showed in section
5 that these 2 observables are {\em definitely} gauge invariant. Accordingly,
Birkhoff's theorem is an "overkill" in the sense that not due care 
has been taken of the functional analysis involved (compare I and 
\cite{24} to make this statement more precise).\\
Birkhoff's theorem refers to the asymptotically flat case only.
In the compact case we neither are able to gauge T and $\Phi$ to zero (note
that for compact topologies there is no difference between gauge and
symmetries) because the observables are manifestly gauge invariant for 
any kind of gauge transformations since surface terms never arise.

\section{Quantum theory}

\subsection{Group theoretical quantization}

We finally come to the quantization of the system. We follow the group
theoretical quantization scheme (see ref. \cite{8}) for each leaf of the
constraint surface separately as discussed at the end of section 5. \\
The configuration space in case II.a has a quite complicated topology
and the discussion of the explicit solution for the corresponding measure
and Hilbert space (except for the case $\gamma=0,\lambda
\le 0$ to which we therefore restrict in the sequel) would by far exceed
the scope of the present paper (the interested reader is referred to
\cite{24}).\\
We then have, referring to \cite{8}, as Hilbert
spaces in case\\
I) $L_2(R^2,dx\wedge dy)$,\\
II.a) 
     $ L_2(R^+\times R,dx/x\wedge dy)$ ,\\
II.b) $L_2(R^2,dx\wedge dy)$.\\
The operators associated to the configuration variables act simply by
multiplication whereas those corresponding to momentum operators are
represented by $\hat{p}=-i\hbar \partial/\partial x$ if the underlying
configuration space is the whole real line and by $\hat{p}=-i\hbar
x\partial/\partial x$ if it is only the positive part of it. They are
obviously self-adjoint with respect to the associated inner products.\\
Over K and for the compact case we so construct 3 different Hilbert
spaces and sets of elementary operators.
We take the 
direct sum of these
Hilbert spaces (thus creating 'sectors'). 
The various elementary operators have then only
diagonal elements. In formulae we have
\beq
\Psi=\left  (  \begin{array}{c}
            \Psi_I \\  \Psi_{II.a} \\  \Psi_{II.b}
               \end{array}
     \right )
\eeq
where we labelled the states that we defined to belong to different sectors
by the associated subscripts and for any operator $\hat{O}$ we have
similarly
\beq
\hat{O}=\left  (  \begin{array}{c}
            \hat{O}_I \\  \hat{O}_{II.a} \\  \hat{O}_{II.b}
               \end{array}
     \right ) \;.
\eeq
Over each asymptotic region, on the other hand, only the Hilbert space 
corresponding to case II.b above is appropriate.

\subsection{The Schroedinger equation}

Let us restrict to the asymptotically flat case and for simplicity that
mass and charge vanish in the interiour compactum K. We thus treat only
the last entry in the decompositions (7.1) and (7.2).\\
The set of elementary observables
is then given by the masses and charges in the various ends and their
conjugate momenta, in other words the Hilbert space is just
$L_2(R^{2n},d^{2n}x)$. We will choose the representation for which the
eigentime and the flux act by multiplication and the mass and the charge by
differentiation. The substitution of classical observables by their quantum
analogues is then unambiguous for the ADM-Hamiltonian and leads to the
following Schroedinger equation
\beq i\hbar\frac{\partial}{\partial t}\Psi(t;\{T_A\},\{\Phi_A\})
     =(-i\hbar\sum_{A=1}^n[N_A(t)\frac{\partial}{\partial T_A}
     +U_A(t)\frac{\partial}{\partial \Phi_A}])\Psi(t;\{T_A\},\{\Phi_A\}) \; .
\eeq
It can be solved trivially by separation :
\beq \Psi(t;\{m_A\},\{\Phi_A\}):=\prod_{A=1}^n\psi_A(t,m_A,\Phi_A) \eeq
and by introducing the functions defined by integrating
\beq \dot{\tau}_A(t):=N_A(t)\; ,\; \dot{\phi}_A(t):=U_A(t) .\eeq
We then find as the general solution
\beq \psi_A(t,T_A,\phi_A)=C_A\exp(k_A\frac{i}{\hbar}[T_A-\tau_A(t)])
                         \times\exp(l_A\frac{i}{\hbar}[\Phi_A-\phi_A(t)])
\eeq
where $C_A$ is a complex number, whereas $k_A,l_A$ must be
real because the spectrum of the momenta, which are self-adjoint, is real.
This set of 2n real numbers labels the state which in particular is an
eigenstate of all momentum operators.
The states are normalizable (to delta distributions) and thus lie in the
completion of the Hilbert-space. These solutions of the time-dependent
Schroedinger equation are obviously peaked at an instant of 'time' t around
the classical solutions (see (6.9)) in the sense that they are strongly 
oscillating off the classical trajectory. The fact that these states look
more like momentum eigenstates than energy eigenstates (for the time
independent Schroedinger equation) is due to the fact that the energy is
linear in all momenta.\\
The quantum-mechanical 'time' happens to coincide with the hypersurface label
t. It is not a Dirac observable, but there are enough candidates for an
intrinsic time in the model, e.g. $T_1$, which {\em is} an observable. Thus
again the so-called Schroedinger time t is not quantized, quantized is
the intrinsic time $T_1$.\\
For compact topologies the states are independent of the Schroedinger time
t. This can be interpreted (see ref. \cite{17}) as the definition of the
Heisenberg picture of quantum theory. Hence, what evolves are not the states
but the observables. We constructed the 'time'-independent Dirac observables.
'Time'-dependent Dirac observables are available along the lines of the 
theory of deparametrization (\cite{5}).

\subsection{Comparison with the operator constraint method}

As for any Yang-Mills theory, the scalar constraint is not any longer
polynomial in the Einstein electric fields (see ref. \cite{18}). One option
how to make sense out of this constraint functional as an operator when
applying the operator constraint method (Dirac method) is to multiply this
constraint by a sufficient power of the electric fields although this leads
to new solutions of the scalar constraint if one does not restrict to
non-degenerate metrics.\\
Let us pursue this recipe for our model. A glance at formula (2.10) reveals
that it is sufficient to multiply the scalar constraint with a factor of
$E^1$ so that in the ordering in which all momenta (for the Ashtekar
polarization) stand to the right, the scalar constraint becomes
\begin{eqnarray}
 & & [(B^2\frac{\delta}{\delta A_2}+B^3\frac{\delta}{\delta A_3})
\frac{\delta^2}{\delta A_1^2}+\frac{1}{2}(B^1\frac{\delta}{\delta A_1}
+\kappa(-\frac{1}{2}\frac{\delta^2}{\delta\omega^2}
+\lambda\frac{\delta^2}{\delta A_1^2}))\times \nonumber\\
 & & (\frac{\delta^2}{\delta A_2^2}+\frac{\delta^2}{\delta A_3^2})]
\Psi[A_I,\omega]=0
\end{eqnarray}
which is a {\em 4th} order functional differential equation. This equation,
of course, does not make any sense the way it stands. First of all, it
involves products of operator-valued distributions and thus should actually
be smeared. Moreover, these distributions are evaluated at the same point
$x\;\in\;\Sigma$ and are thus meaningless, in general, unless one regularizes
them. Finally, consistency of the Dirac method (see ref. \cite{19}) requires
that all constraints form a closed operator subalgebra. A formal calculation
of the commutators of the constraints in which all momenta are to the right
reveals that they do not close in the sense that the commutator is
proportional to a constraint operator, however the constant of
proportionality depends on the fields and stands to the right. A trial and
error procedure in order to find a correct ordering did not succeed. However,
the way the constraints were solved classically suggests how to solve
the quantum constraints : by a suitable choice of polarization. In the
following we will only treat the case II.b for one asymptotical end or
for the compact case and we denote this region by M.\\
We simply choose the following set of canonical pairs (compare (4.3))
\beq (p,-\omega;\pi_\gamma,-\gamma;\pi_1,-B^1;\alpha,{\cal G}) \eeq
and choose the representation in which state functionals depend only on
$p,\pi_\gamma,\pi_1\;\mbox{and}\;\alpha$. The leaf of the constraint
surface corresponding to case II.b is defined (in these coordinates) by
\begin{eqnarray}
0 & = & {\cal G} \\
0 & = & p' \\
0 & = & \alpha'{\cal G}+\pi'_\gamma(-\gamma)-(-\pi_1)(B^1)' \\
0 & = & \pi_\gamma'(\pi_\gamma B^1+\kappa(p^2/2+\lambda\pi_\gamma^2))
       -(\pi_\gamma B^1+\kappa(p^2/2+\lambda\pi_\gamma^2)){\cal G}
       +2\pi_\gamma^2 (B^1)'
\end{eqnarray}
where one obtains (7.11) and (7.12) most conveniently as follows :
one writes formulas (4.6) in the above coordinates and obtains 2 equations
\begin{eqnarray}
C_1 & = & \sqrt{E}(\pi_\gamma B^1+\kappa(p^2/2+\lambda\pi_\gamma^2))
\cos(\beta)+2\pi_\gamma^2(\gamma\sqrt{A}\cos(\alpha)+\frac{(B^1)'}{\sqrt{A}}
\sin(\alpha))  \nonumber \\
C_2 & = & \sqrt{E}(\pi_\gamma B^1+\kappa(p^2/2+\lambda\pi_\gamma^2))
\sin(\beta)+2\pi_\gamma^2(\gamma\sqrt{A}\sin(\alpha)-\frac{(B^1)'}{\sqrt{A}}
\cos(\alpha))
\end{eqnarray}
of which one takes the following combinations
\beq \frac{1}{\sqrt{A}}(\cos(\alpha)C_1+\sin(\alpha)C_2)\;\mbox{and}\;
     \sqrt{A}(\sin(\alpha)C_1-\cos(\alpha)C_2)
\eeq
which lead directly to (7.11) and (7.12).\\
Now in the polarization chosen, the constraints are linear in momenta except
for the term $\propto {\cal G}$ (which could also be dropped alternatively
without losing a constraint) in (7.12).
However, the coordinate $\alpha$
does nowhere appear in our constraints such that we can simply treat
these momenta as C numbers when computing commutators. Since, as proved
in section 5.3, the constraints
on the leaf also close and since they are linear in the momenta $\gamma\;
\mbox{and}\;B^1$ it follows that they close also as operators irrespective
of the ordering chosen. For quantum theory, the ordering that they stand
to the right is the most useful one and thus we take (7.9)-(7.12) as the
ordering for quantum theory with the substitutions
\beq
{\cal G}\rightarrow\kappa\frac{\delta}{\delta\alpha},\;
\omega\rightarrow i\frac{\delta}{\delta p},\;
B^1\rightarrow -\kappa\frac{\delta}{\delta\pi_1},\;
\gamma\rightarrow -\kappa\frac{\delta}{\delta\pi_\gamma} \; .
\eeq
Let us then solve the quantum constraints. (7.9) and (7.10) applied to a
functional of the configuration variables and set equal to zero reduces
the dependence of a physical state on these variables to the form
\beq \tilde{\Psi}[p,\alpha,\pi_1,\pi_\gamma]=
\delta[p']\Psi(p;\pi_1,\pi_\gamma] \;
\eeq
where the notation means that $\Psi$ depends on the {\em functions} $\pi_1,
\pi_\gamma$ but on the {\em parameter} p.\\
The dependence of a physical state on the variable $\pi_1$ is determined
by (7.12) : multiplying this constraint by $\pi_\gamma^{-3/2}$ {\em from
the left} one obtains
\beq  [\frac{\kappa}{\sqrt{\pi_\gamma}}(-\pi_\gamma\frac{\delta\Psi}{\delta
\pi_1}
+\frac{\lambda}{3}\pi_\gamma^2\Psi-\frac{1}{2}
p^2\Psi)]'=0 \;.\eeq
This equation says that the bracket term is an arbitrary spatial constant P.
It equation has the general solution
\beq \Psi=\exp(\int_M dx \pi_1(\frac{\lambda}{3}
\pi_\gamma-\frac{p^2}{2\pi_\gamma}))\times
\psi(p,\int_M dx\frac{\pi_1}{\sqrt{\pi_\gamma}});\pi_\gamma]
=:\exp(\theta(p))\psi(p,T;\pi_\gamma]  \eeq
where we recognize the expression for the eigentime T in the first argument
of $\psi_\pm$ and there is still a functional dependence on $\pi_\gamma$.
Finally, we obtain from the remaining constraint (7.17)
\begin{eqnarray}
0 & = & [\pi_1(\frac{\delta}{\delta\pi_1})'
-\pi_\gamma'\frac{\delta}{\delta\pi_\gamma}]\Psi \nonumber\\
& = & -\pi_\gamma'
\exp(\theta)\frac{\delta\psi}{\delta\pi_\gamma}
\end{eqnarray}
whence $\psi$ is an ordinary function of T only, i.e.
\beq \psi:=\psi(p,T)\; . \eeq
Thus, the nontrivial information about the quantum state $\tilde{\Psi}$
is contained in the function $\psi(p,T)$ and therefore we choose the measure
for the inner product to be 
\[ d\mu:=\exp(-\theta)dp\wedge dT \]
which turns the conjugate momenta $-\Phi, m$ into self-adjoint operators.
We therefore have established that the quantum theories as obtained via 
either the reduced phase space approach or the operator constraint approach
can be made equivalent in this case by going to the appropriate quantum
representation.

\section{Conclusions}

Let us summarize the new results of the present paper :\\
\\
$\bullet$
We showed that the reduced phase space method can be applied to spherically
symmetric Einstein-Maxwell theory to complete the full quantization
programme with full mathematical rigour.\\
$\bullet$
The analysis was carried out in the Ashtekar variables rather than in the
geometrodynamical (ADM) variables. This is the first model for (3+1) 
gravity coupled to gauge fields that has been quantized completely
(in the Ashtekar variables).\\
$\bullet$
The reduced phase space is coordinatized completely, in every
asymptotical end, by the gravitational mass, the eigentime, as well as the
electric charge and the (formal) magnetic flux as measured by an asymptotic
observer in that end of the time slice of the underlying 4-manifold. These
interpretations of the Dirac observables hold "on shell" of the reduced 
dynamical phase space, in the asymptotically flat case. These observables
are never mentioned in the textbook treatments of the Reissner-Nordstr{\o}m
solution \cite{15} since no due care is taken of the distinction between
gauge and symmetry.\\
$\bullet$
In the compact case we were able to treat also the case of a nonvanishing
cosmological constant.\\
$\bullet$
There are several "sectors" of the theory due to the fact that the 
scalar constraint is nonlinear in the momenta. These sectors are 
carefully treated using group theoretical and algebraic quantization
techniques.\\
$\bullet$
Classically we were able to compute the Hamilton-Jacobi functional, in
quantum theory we succeeded in solving the Schroedinger equation.\\  
$\bullet$
Besides performing the quantization via symplectic techniques, we were also
able to complete the quantization using the operator constraint (Dirac)
method. The resulting quantum theories obtained turn out to be equivalent.\\
$\bullet$
Perhaps the most interesting technical result of the present paper is the
successful quantization of a model for (3+1) gravity coupled to an
abelian gauge field whose constraints are fourth order in the momenta.\\
\\
The conclusions that we may draw are as follows :\\
\\
$\bullet$
We found it convenient to work out the symplectic reduction of the model
not in the Ashtekar polarization but in a polarization which mixes both
the triad and the connection coefficients although, as usual, it turned out
that it is convenient to start with the Ashtekar polarization since it
simplifies the analysis tremendously. Of course, after the difficult part
of the work is performed, i.e. to find the observables, one can return to
the Ashtekar polarization : one has simply to solve equation (4.12) for $E^1$
in terms of $B^1$, p and $\lambda$ and plug this into the expressions for
the observables. For nonvanishing cosmological constant, eq. (4.12) and the
corresponding formulas are somewhat lengthy and one has 4 roots which have
to be reconciled with the reality of $E^1$. Let us restrict therefore to the
more feasible and physically more relevant case $\lambda=0$. Now, one only has
to solve a  quadratic equation. The 2 roots are
\beq E^1=\frac{1}{2(B^1)^2}([p^2\kappa B^1+m^2]\pm\sqrt{[p^2
\kappa B^1+m^2]^2-[p^2\kappa B^1]^2}) \;, \eeq
but only the one with the positive sign is physical since in the case of
vanishing charge we have to recover the result in I (the negative sign leads
to $E^1=0$). Note that the reality of $E^1$ requires that
$B^1\le-m^2/(2p^2\kappa)$, however, this imposes no further constraints on the 
range of $T,\;\Phi$ because 
this yields for the integrand of the variables T and
$\Phi+\int_\Sigma dx \omega$ respectively
\beq
-2\frac{A_1+[\arctan(\frac{A_3}{A_2})]'}{(2B^1)^{2-n}}\frac{[p^2\kappa+
\frac{m^2}{B^1}+\sqrt{[p^2\kappa+\frac{m^2}{B^1}]^2-[p^2\kappa]^2}]^{2-n}}
{p^2\kappa+1/2(\frac{m^2}{B^1}+\sqrt{[p^2\kappa+\frac{m^2}{B^1}]^2
-[p^2\kappa]^2})}
\eeq
where $n=1/2$ and $n=1$ respectively and $B^1=1/2((A_2)^2+(A_3)^2-2)$.
This expression is much more complicated than the one in (4.21) in terms
of $\pi_1\;\mbox{and}\;\pi_\gamma$ and it is thus
suggested that in general the polarization that one starts with will not
turn out to be the natural one for the problem at hand. \\
$\bullet$
Formula (8.2) also enables one to straightforwardly rewrite all the results
given in the present paper in terms of the geometrodynamical 
variables along the lines of the procedure given in the first
reference of \cite{1}. After all, the Ashtekar variables differ from the 
ADM variables merely by a canonical transformation. However, as the
reader may check by himself, the computations are rather tedious and
lengthy and the formulas become less feasible than when using the Ashtekar 
variables.\\
$\bullet$
It should be stressed that the observables found have {\em genuine} volume
integral representations and therefore they do not vanish even in the compact
case. Furthermore, the matter-coupling suggests that the T-variable should
be considered on equal footing with the $\Phi$-variable. Since the latter also
occurs in (1+1) canonical Maxwell-theory in a Minkowski background geometry,
the very existence of T, against
which various arguments were raised 
in the past, e.g. that it always should be possible to gauge it to
zero, that spherically symmetric gravitational fields have no proper
reduced phase space etc., should appear less unnatural. \\
\\
\\
{\large Acknowledgements} \\ 

I thank foremost Prof. H.A. Kastrup for many stimulating
interactions and Prof. A. Ashtekar and 
Prof. L. Smolin and the Syracuse relativity group for fruitful discussions.
This project was supported by the Graduiertenkolleg of the Deutsche 
Forschungsgemeinschaft.

\begin{appendix}

\section{Computations related to section 6}

We give here the explicit calculations necessary to prove the gauge
invariance of the observables, their evolution laws and that they satisfy
a canonical Poisson algebra. In order to display the fine tuning referred to
earlier, let us compute the Poisson bracket between the contribution to the
observables, that vanishes or does not vanish on the constraint surface
respectively, with the symmetry generators, separately.\\
\\
In the following calculations we will make frequent use of the constraint
equations $(E^1)'=g(B^1)',\; E^2=g B^2\;\mbox{and} E^3=g B^3\;\mbox{where}\;
g:=-2 (E^1)^2/(B^1 E^1+\kappa(\lambda (E^1)^2+p^2/2))$, (see (4.12)),
which is allowed since we are working on the constraint
surface. What we have to do is simply to insert the variations of the basic
variables (6.1) in the expressions for the variation of the observables
(6.3).\\
Let us start with the variation of m under a general symmetry
transformation (recall the definition of G above eq. (6.1)) :
\begin{eqnarray}
\frac{1}{i\kappa}\delta m & = & \frac{1}{i\kappa}\int_\Sigma dx D
\sqrt{E^1}[-g^{-1}\delta E^1+
(A_2\delta A_2+A_3\delta A_3+\frac{\kappa p}{2(E^1)^2}\delta p]\nonumber \\
 & = & \int_\Sigma dx D\sqrt{E^1}[-g^{-1}(L(E^1)'-ME^1(A_2 E^2+A_3 E^3))
\nonumber\\
& & +(A_2(-\Lambda A_3+LA_2'+M(E^1 B^2-E^2(B^1+\kappa(p^2/(2E^1)+\lambda E^1)))
\nonumber\\
& & +A_3(\Lambda A_2+LA_3'+M(E^1 B^3+E^3(B^1+\kappa(p^2/(2E^1)+\lambda E^1)))
+\frac{\kappa p}{2(E^1)^2}\delta p]\nonumber \\
& = & \int_\Sigma dx D\sqrt{E^1}[-g^{-1}(L (E^1)'+ME^1(A_2 E^2+A_3 E^3))
+L (B^1)'+M E^1\gamma A] \nonumber\\
& = & \int_\Sigma dx D\sqrt{E^1}L[-g^{-1}(E^1)'+(B^1)']=0 \;.
\end{eqnarray}
Hence, m is a constant of motion, not only gauge invariant.\\
The equation for p is trivial, since it is simultaneously a basic variable :
\beq \delta p=0 \; . \eeq
The equation for T is much more complicated. We begin with the variation of
the non-constraint part of $\delta T$:
\begin{eqnarray}
&  & i\kappa\int_{\Sigma}dx[\frac{g}{\sqrt{E^1}}\delta A_1
+(\frac{g^2\gamma}{2\sqrt{E^1}^3}A_2+(\frac{g}{\sqrt{E^1}})'\frac{A_3}{A})
\delta A_2 \nonumber \\
 & & +(\frac{g^2\gamma}{2\sqrt{E^1}^3}A_3-(\frac{g}{\sqrt{E^1}})'
 \frac{A_2}{A})\delta A_3
+\frac{\gamma}{\sqrt{E^1}^3}(3/2 g+\frac{g^2}{2 E^1}(B^1+2/3\lambda
\kappa E^1))\delta E^1] \nonumber \\
 & = &\int_{\Sigma}dx[\frac{g}{\sqrt{E^1}}(-\Lambda'+(LA_1)'
 +M(B^2 E^2+B^3 E^3)+E\kappa/2(-p^2/(2(E^1)^2)\nonumber\\
& & +\lambda))) \nonumber\\
& &+(\frac{g^2\gamma}{2\sqrt{E^1}^3}A_2+(\frac{g}{\sqrt{E^1}})'\frac{A_3}{A})
(-\Lambda A_3+LA_2'+M(E^1 B^2+E^2(B^1+\kappa(p^2/(2E^1)\nonumber\\
 & & +\lambda E^1)))\nonumber \\
& &+(\frac{g^2\gamma}{2\sqrt{E^1}^3}A_3-(\frac{g}{\sqrt{E^1}})'\frac{A_2}{A})
(\Lambda A_2+LA_3'+M(E^1 B^3+E^3(B^1+\kappa(p^2/(2E^1)\nonumber\\
& & +\lambda E^1))) \nonumber \\
& & +\gamma\frac{\partial}{\partial E^1}(\frac{g}{\sqrt{E^1}})
(L (E^1)'-ME^1(A_2 E^2+A_3 E^3))] \nonumber \\
& = &\int_{\Sigma}dx[\frac{g}{\sqrt{E^1}}(-\Lambda'+(LA_1)'
 +M E(1/g+\kappa/2(-p^2/(2(E^1)^2)+\lambda)) \nonumber\\
& &+(\frac{g^2\gamma}{2\sqrt{E^1}^3}(L(B^1)'-M E^1\gamma A)
+(\frac{g}{\sqrt{E^1}})'\frac{1}{A}(-\Lambda A-L\alpha' A\nonumber\\
 & & +M(E^1(B^1)'-1/g(A_3 E^2-A_2 E^3)2 E^1) \nonumber\\
& & +\gamma\frac{\partial}{\partial E^1}(\frac{g}{\sqrt{E^1}})
(L (E^1)'-M g E^1 A\gamma)]
\end{eqnarray}
where we used the identity $A_2 B^2+A_3 B^3=A\gamma$. We observe that the
2 first terms in the first line of the last equality can be combined with 2
companions in the 2nd bracket of the 2nd line to obtain a total spatial
derivative and we use the Gauss constraint in the same bracket to simplify
this bracket further. We thus obtain when using $\partial g/\partial B^1
=g^2/(2 E^1)$ for the non-constraint part of $\delta T$
\begin{eqnarray}
& &\int_{\Sigma}dx[(\frac{g}{\sqrt{E^1}}(LA_1-
\Lambda))'+\frac{g}{\sqrt{E^1}}M E(1/g+\kappa/2(-p^2/(2(E^1)^2)+\lambda))
\nonumber\\
& &+(\frac{g^2\gamma}{2\sqrt{E^1}^3}(L(B^1)'-M E^1\gamma A)
-(\frac{g}{\sqrt{E^1}})'(L\gamma A+M E^1(B^1)') \nonumber\\
& & +\gamma\frac{\partial}{\partial E^1}(\frac{g}{\sqrt{E^1}})
(L (E^1)'-M g E^1 A\gamma)] \nonumber\\
& = &\int_{\Sigma}dx[(\frac{g}{\sqrt{E^1}}(LA_1-\Lambda))'
+\frac{g}{\sqrt{E^1}}M E(1/g+\kappa/2(-p^2/(2(E^1)^2)+\lambda)) \nonumber\\
& &+(\frac{g^2\gamma}{2\sqrt{E^1}^3}(L(B^1)'-M E^1\gamma A)
-(\frac{\partial}{\partial E^1}(\frac{g}{\sqrt{E^1}})(E^1)'
 +\frac{\partial}{\partial B^1}(\frac{g}{\sqrt{E^1}})(B^1)')\times
\nonumber\\
& & (L\gamma A+\frac{M}{A} E^1(B^1)') 
+\gamma\frac{\partial}{\partial E^1}(\frac{g}{\sqrt{E^1}})
(L (E^1)'-M g E^1 A\gamma)] \nonumber\\
& = &\int_{\Sigma}dx[(\frac{g}{\sqrt{E^1}}(LA_1-\Lambda))'
+\frac{g}{\sqrt{E^1}}M E(1/g+\kappa/2(-p^2/(2(E^1)^2)+\lambda)) \nonumber\\
& &-(\frac{g^2\gamma}{2\sqrt{E^1}^3}M E^1\gamma A)
-(\frac{\partial}{\partial E^1}(\frac{g}{\sqrt{E^1}})(E^1)'
 +\frac{\partial}{\partial B^1}(\frac{g}{\sqrt{E^1}})(B^1)')
 (\frac{M}{A} E^1(B^1)') \nonumber\\
& & -\gamma\frac{\partial}{\partial E^1}(\frac{g}{\sqrt{E^1}})
    M g E^1 A\gamma)] \nonumber\\
& = &\int_{\Sigma}dx[(\frac{g}{\sqrt{E^1}}(LA_1-\Lambda))'
+\frac{g}{\sqrt{E^1}}M E(1/g+\kappa/2(-p^2/(2(E^1)^2)+\lambda)) \nonumber\\
& & -(A\gamma^2+\frac{1}{A}((B^1)')^2)(\frac{g^2}{2\sqrt{E^1}^3}M E^1
     +\frac{\partial}{\partial E^1}(\frac{g}{\sqrt{E^1}})M g E^1)] \; .
\end{eqnarray}
We now use the identity $A\gamma^2+((B^1)')^2/A=(B^2)^2+(B^3)^2=E/(g^2)$
and can collect all non-boundary terms with a common prefactor $E$ in the last
equation :
\begin{eqnarray}
 &  &\int_{\Sigma}dx[(\frac{g}{\sqrt{E^1}}(LA_1
-\Lambda))'
+M E (\frac{g}{\sqrt{E^1}}(1/g+\kappa/2(-p^2/(2(E^1)^2)+\lambda) \nonumber\\
& & -(\frac{1}{2\sqrt{E^1}} 
     +\frac{E^1}{g}\frac{\partial}{\partial E^1}(\frac{g}{\sqrt{E^1}})))] \; .
\end{eqnarray}
Now, one has only to use $\partial/\partial E^1(\frac{g}{\sqrt{E^1}})=
g^2/(4(E^1)^{7/2})(-(B^1 E^1+\kappa p^2)+\kappa(\lambda (E^1)^2-p^2/2))$ in
order to show that the last equation becomes
\beq \int_{\partial\Sigma}\frac{g}{\sqrt{E^1}}
(LA_1-\Lambda)\; . \eeq
As for the constraint part of $\delta T$, we can, as already derived 
in section 6, refer to formula 6.2 and obtain on the constraint surface
\begin{eqnarray}
 & & - \delta G[A_1\frac{g\sqrt{2} E^1 E^2}{(\sqrt{E^1})^3 E A},
 \frac{g\sqrt{2} E^1 E^2}{(\sqrt{E^1})^3 E A},
\frac{g\sqrt{2}E^3}{(\sqrt{E^1})^3 E A},0] \nonumber \\
& = & -\int_{\partial\Sigma}[(L A_1-\Lambda)\frac{g\sqrt{2} E^1 E^2}
{(\sqrt{E^1})^3E A})(A_2 E^3-A_3 E^2)
\nonumber \\
& & +(M A_1\frac{g\sqrt{2} E^1 E^2}{(\sqrt{E^1})^3 E A}
-\Lambda\frac{g\sqrt{2}E^3}{(\sqrt{E^1})^3 E A}) (A_2 E^2+A_3 E^3)E^1
\nonumber\\
& &+(L \frac{g\sqrt{2}E^3}{(\sqrt{E^1})^3 E A}-M\frac{g\sqrt{2} E^1 E^2}
{(\sqrt{E^1})^3 E A})\times \nonumber\\
& & (A_1 E^1(A_2 E^2+A_3 E^3)+\frac{1}{2}E(B^1+\kappa(p^2/(2 E^1)+\lambda 
E^1))]
\; .
\end{eqnarray}
We use the boundary conditions for symmetry transformations to simplify this
expression of which most terms vanish at infinity. The second term is
already $O(1/x)$. The first summand in the first bracket in the last term
is $O(1/x^2)$ the second is $O(1/x)$, while in the second bracket of the
last term the first summand is $O(1)$ and the second has a $O(x)$ term in
leading order (since we are dealing with an asymptotic region we have to set
the cosmological constant equal to zero). Finally, using the Gauss constraint
in the first term we see that it is $O(1)$ altogether so that we end up
with the following contribution of the constraint part of $\delta T$ :
\begin{eqnarray}
&  & -\int_{\partial\Sigma}[(L A_1-\Lambda)\frac{g\sqrt{2} E^1 E^2}
{(\sqrt{E^1})^3E A})(E^1)'
\nonumber \\
& &+M\frac{g\sqrt{2} E^1 E^2}{(\sqrt{E^1})^3 E A}\frac{1}{2}E B^1] \; .
\end{eqnarray}
Now, since $L A_1-\Lambda=O(1/x^2),A\rightarrow 2,E^1\rightarrow x^2,(E^1)'
\rightarrow 2x,g=O(x^3),
\sqrt{2}E^2\rightarrow 2x,E\rightarrow 2x^2$ (plus higher orders respectively),
we can replace the first term in the last formula by
\beq - \int_{\partial\Sigma}(L A_1-\Lambda)\frac{g}{\sqrt{E^1}} \eeq
which {\em exactly} cancels the contribution (A.6) from the part of $\delta T$
which does not vanish on the constraint surface.\\
Finally, since $g\rightarrow -2 x^3/(\sqrt{2}a_3)=2^3/m,M x^2=O(1),B^1
\rightarrow -m/x$,
we can conclude that the complete variation of T is indeed given by 
\beq 
\delta T=\lim_{x\to +\infty}\int_{\partial\Sigma}(Mx^2) 
\eeq
which is just the lapse at infinity for a symmetry transformation while
T is indeed gauge invariant for a gauge transformation.\\
The variation of $\Phi$ is now fairly easy to derive : it is nearly the
same as that for T except that one has to divide all expressions
by one more power of $\sqrt{E^1}$. The total differentials that arise
therefore all vanish except for that coming from the variation of $\omega$ :
\beq \delta -\int_\Sigma dx\omega=+\int_{\partial\Sigma}U
-\frac{\kappa p}{2}\int_\Sigma M E/E^1 \;.\eeq
The second term in the last equation is due to the following identity
\beq \sqrt{E^1}(\frac{E^1}{g}\frac{\partial}{\partial E^1}({g}{E^1})+\frac{1}{2E^1})
=\frac{E^1}{g}\frac{\partial}{\partial E^1}({g}{\sqrt{E^1}}) \eeq
which shows that the changes in the term $\propto E$ due to the power
change of $\sqrt{E^1}$ and the appearance of the $\omega$ in $\Phi$ as
compared to T exactly cancel each other so that in conclusion
\beq \delta\Phi=+\lim_{x\to +\infty}\int_{\partial\Sigma} U \; .\eeq
We have thus arrived at the expected transformation laws for our reduced
variables as claimed in section 6.\\
It should be clear by now how to prove that they form a closed canonical
Poisson algebra in which (T,m) and ($\Phi$,p) are canonical pairs. The
computations are rather tedious to perform and do not give further insight
into the ideas involved, so we refrain from displaying them here.

\end{appendix}

\end{document}